# Theoretical analysis of Fresnel reflection and transmission in the presence of gain media

Masud Mansuripur† and Per K. Jakobsen‡

†College of Optical Sciences, The University of Arizona, Tucson, Arizona, USA

‡Department of Mathematics and Statistics, UIT The Arctic University of Norway, Tromsø, Norway

**Abstract**. When a monochromatic electromagnetic plane-wave arrives at the flat interface between its transparent host (i.e., the incidence medium) and an amplifying (or gainy) second medium, the incident beam splits into a reflected wave and a transmitted wave. In general, there is a sign ambiguity in connection with the $k$-vector of the transmitted beam, which requires at the outset that one decide whether the transmitted beam should grow or decay as it recedes from the interface. The question has been posed and addressed most prominently in the context of incidence at large angles from a dielectric medium of high refractive index onto a gain medium of lower refractive index. Here, the relevant sign of the transmitted $k$-vector determines whether the evanescent-like waves within the gain medium exponentially grow or decay away from the interface. We examine this and related problems in a more general setting, where the incident beam is taken to be a finite-duration wavepacket whose footprint in the interfacial plane has a finite width. Cases of reflection from and transmission through a gainy slab of finite-thickness as well as those associated with a semi-infinite gain medium will be considered. The broadness of the spatiotemporal spectrum of our incident wavepacket demands that we develop a general strategy for deciding the signs of all the $k$-vectors that enter the gain medium. Such a strategy emerges from a consideration of the causality constraint that is naturally imposed on both the reflected and transmitted wavepackets.

**1. Introduction**. The well-known Fresnel reflection and transmission formulas describe the behavior of a monochromatic plane-wave of frequency $\omega$, incident on a flat interface between two homogenous media.[1] The geometry is shown in Fig.1(a), in which media 1 and 2 are both semi-infinite, and we call this the "single-surface" problem. We will also consider the "finite-slab" problem shown in Fig.1(b), in which a finite thickness medium 2 is sandwiched between semi-infinite media 1 and 3. To calculate the response of the system by the usual Fresnel method, we assume the existence of two counter-propagating plane-waves in each medium (one for each wave-vector $\boldsymbol{k}$ allowed by Maxwell's equations) except for the final transmission medium, for which we handpick one of the two $k$-vectors on the basis of some "commonsense" argument such as "energy should flow to the right" or "the field amplitude must decay as $z \to \infty$." Then, we enforce the boundary conditions at each interface to unambiguously determine the amplitude of each plane-wave. Next, to determine the response of the system to a more realistic stimulus, such as a beam or a pulse, we decompose the stimulus into plane-waves by Fourier transformation, apply the Fresnel response in the Fourier domain, and then recompose the resulting waves by an inverse Fourier transformation. With these two steps, (i) determining the response to a single $(\boldsymbol{k}, \omega)$ plane-wave, and (ii) treating the incident stimulus as a superposition of such plane-waves, the problem is solved in generality, or so the conventional wisdom goes.[1-3]

However, if the transmission medium has gain (as opposed to being lossy or lossless), both steps described above become problematic. First of all, in the single-surface problem, there is no longer a commonsense argument that the field amplitude must decay as $z \to \infty$, leading to an ambiguity in the choice of the transmitted wave-vector, and resulting in uncertainty about the Fresnel response. Secondly, and more surreptitiously, the usual Fourier transformation methods break down, because the gain medium can cause the system response function to be non-analytic in the upper-half of the complex frequency plane, $\omega$. By considering only the real $\omega$-axis, as is usually done, and not accounting for the imaginary part of $\omega$, these calculations can result in reflected and transmitted pulses that violate causality. (See reference [4] for the behavior that results when the problem is treated in this naïve, but common, way.)

The ambiguity in the direction of the transmitted wave-vector has been discussed most often in the context of total internal reflection (TIR) at the interface between a transparent dielectric and a gain medium.[4-6] TIR is a special case of the single-surface problem, where the real-valued $\varepsilon_1$ is greater than the real-valued $\varepsilon_2$, and a monochromatic plane-wave of frequency $\omega$ arrives at the incidence angle $\theta_{\text{inc}}$ that exceeds the critical TIR angle $\theta_{\text{TIR}} = \sin^{-1}\left(\sqrt{\varepsilon_2/\varepsilon_1}\right)$. In this case, given the dispersion relation that yields the $z$-component of the transmitted $k$-vector in terms of $k_x$, $\omega$, and the speed of light in free space, $c$, as $k_{2z} = \pm(\omega/c)\sqrt{\varepsilon_2 - (ck_x/\omega)^2}$, the quantity under the radical becomes negative, resulting in an evanescent wave that could either decay or grow (exponentially) as it recedes from the interface.

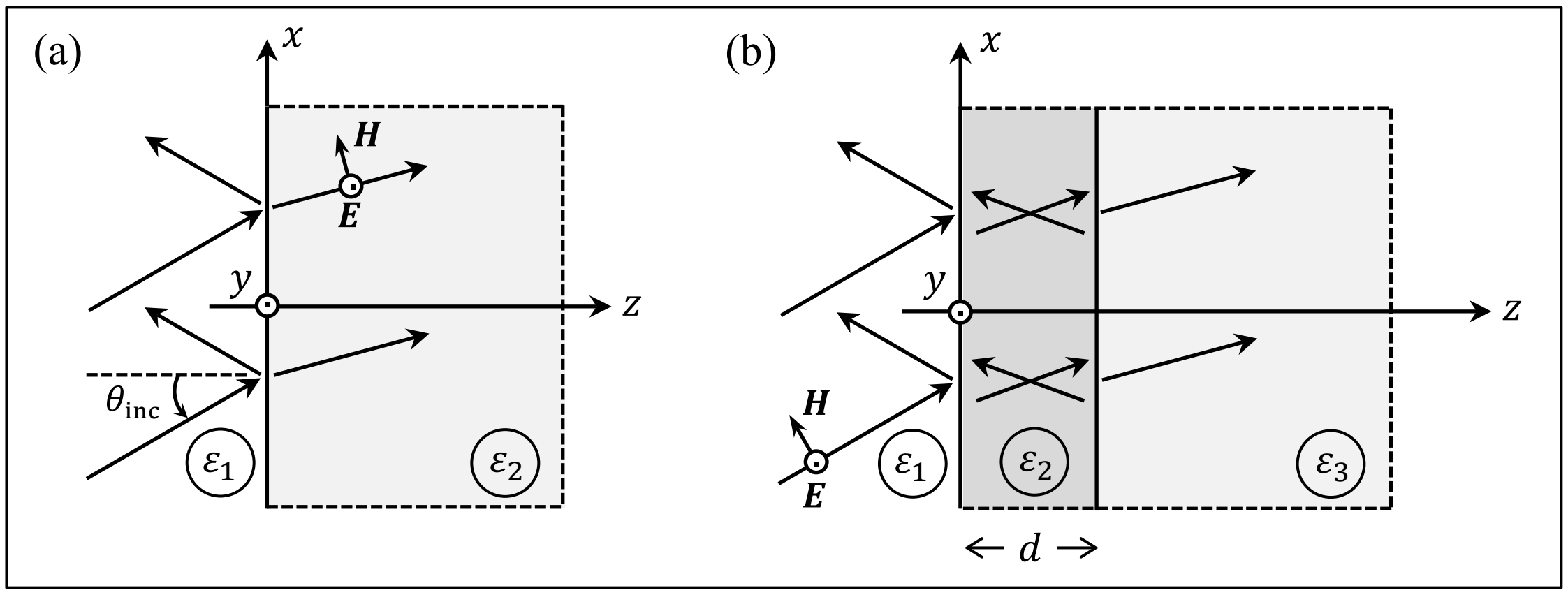

**Fig.1**. A finite-duration and finite-width wavepacket arrives at the flat interface between a semi-transparent (i.e., weakly absorptive) medium of permittivity $\varepsilon_1(\omega)$ and a gain medium of permittivity $\varepsilon_2(\omega)$. The interfacial $xy$-plane is located at $z = 0$. (a) In the "single-surface" problem, the gain medium is semi-infinite. (b) In the "finite-slab" problem, the gain medium of thickness $d$ is sandwiched between the incidence medium and another semi-transparent medium of permittivity $\varepsilon_3(\omega)$. In general, when a material medium has permeability $\mu(\omega) = 1$, its refractive index $n(\omega)$ equals the square root of its permittivity $\varepsilon(\omega)$. The analysis in this paper is restricted to $s$-polarized waves (in which the electric field is parallel to the $y$-axis), though the method is equally applicable to $p$-polarized waves as well.

When the transmission medium is lossless, the commonsense argument compels us to choose the exponentially decaying evanescent wave.[1-3] If the transmission medium happens to be weakly absorptive, its complex dielectric constant $\varepsilon_2' + \mathrm{i}\varepsilon_2''$ brings about a small reduction in the Fresnel reflection coefficient at the interface by directing a fraction of the incident energy into medium 2, which this medium subsequently absorbs. The general characteristics of the quasi-evanescent wave, however, will *not* change drastically compared to the transparent case, provided that the imaginary part $\varepsilon_2''$ of $\varepsilon_2$ is reasonably small. Specifically, the transmitted wave continues to exponentially decay away from the interface, albeit with a small (non-zero) component of its Poynting vector directing the electromagnetic (EM) energy along the propagation direction, which we have taken to be the $z$-axis. In contrast, if the transmission medium has gain, which occurs when $\varepsilon_2''$ assumes a negative value at the frequency $\omega_{\text{inc}}$ of the incident (monochromatic) plane-wave, there is no simple commonsense argument to decide whether the EM wave within medium 2 should grow or decay exponentially along the $z$-axis.[4-6] (This question of TIR from a gain medium is not merely an academic curiosity, but one that has important practical applications and consequences in the context of fiber-optical lasers and amplifiers in which the core is passive and the cladding constitutes the gain medium.[7])

In this paper, we explain how to solve the Fresnel problem without having to handpick a particular wave-vector $\boldsymbol{k}$ in the transmission medium. The key is to analyze the system's response not just at a single frequency, but in the entire complex frequency plane, $\omega = \omega' + \mathrm{i}\omega''$, and impose the requirement that the system obey causality. In this way, the correct choice for the direction of the wave-vector at every value of $\omega$ naturally emerges, eliminating the ambiguity that previously required the choice of one solution over another. The causality constraint thus replaces the requirement for "commonsense," and the method works just as well for gainy media as it does for transparent and lossy media.

By choosing an incident wavepacket that has a finite duration as well as a finite footprint at the interface between media 1 and 2, we broaden the scope of the investigation to include a continuum of temporal frequencies $\omega$ as well as spatial frequencies $k_x$ within the spectral profile of the incident wave. As it turns out, answering the narrow question of how to solve the Fresnel problem for a single incident plane-wave, $(k_x, \omega)_{\text{inc}}$, requires a comprehensive solution for the entire spatiotemporal spectrum of the incident wavepacket. This is true not only for the single-surface problem depicted in Fig.1(a), but also for the finite-slab problem of Fig.1(b). In the special case of TIR from a semi-infinite gain medium, the correct $k_{2z}$ will be seen to carry the plus sign for some values of $(k_x, \omega)$ and the minus sign for the others. Thus, the solution to this problem is not as simple as a prescription for which sign to choose below, and which sign above, the critical angle $\theta_{\text{TIR}}$. Instead, the problem leads us to consider the entire complex $k_x$-plane and, along the way, it redefines our notion of what it means to "cross the critical angle." The problem also illuminates important details about spatiotemporal spectral composition: the plane-waves that constitute a wavepacket must be carefully chosen to avoid violating causality.

The connection between a system's frequency response and causality is taught at the undergraduate level, typically in the context of the Kramers-Kronig relations.[2] In that case, the system under consideration (e.g., an individual dipole) is driven by a time-dependent stimulus, the response to which unfolds as a function of time only. When the system's stimulus and response happen to be functions of both space and time, as in the Fresnel problem, there are subtleties that require careful consideration. In what follows, we address these issues in the general framework of linear response theory before proceeding to apply the concepts to the Fresnel problem.

The organization of the paper is as follows. In the next section, we examine the connection between causality and the complex-plane representation of spatiotemporal frequencies $(\boldsymbol{k}, \omega)$ in a linear system. An examination of the analyticity of the system's transfer function in the entire complex plane reveals the circumstances under which the superposition integrals involving the $k_x$ variable are best carried out in the complex $k_x$-plane along a contour that deviates from the real $k_x'$-axis. Then, in Sec.3, we formulate the Fresnel problem in the presence of gainy media as one such linear system to which the aforementioned complex-plane techniques apply. In the sections that follow, we dissect the Fresnel problem's transfer function and proceed to examine it one piece at a time.

Section 4 is devoted to a description of the Lorentz oscillators that underlie the frequency-dependent dielectric functions $\varepsilon(\omega)$ of the material media — be they the nearly-transparent incidence and transmittance media 1 and 3, or the amplifying gain medium 2. In Sec.5, we describe the Fresnel reflection and transmission coefficients for the two systems under consideration, namely, the system depicted in Fig.1(a) involving a semi-infinite gain medium, and the system of Fig.1(b), where a finite-thickness gainy slab is sandwiched between two nearly-transparent dielectric media. The geometric configuration of these systems is such that the $k$-vectors in each medium will have a component $k_x$ along the $x$-axis and a component $k_z$ along the $z$-axis. Given an incident plane-wave with $(\boldsymbol{k}, \omega) = (k_x\hat{\boldsymbol{x}} + k_{1z}\hat{\boldsymbol{z}}, \omega)$, Maxwell's boundary conditions obligate the excited plane-waves in media 1, 2, and 3 to have the same $k_x$ value as the incident wave; this shared value of $k_x$ will be treated as a complex entity and written as $k_x' + \mathrm{i}k_x''$. The corresponding $k_z$ in each medium is subsequently determined from the dispersion relation, $k_x^2 + k_z^2 = (\omega/c)^2\varepsilon(\omega)$, where $c$ is the speed of light in vacuum. The dependence of $k_z$ on $\varepsilon(\omega)$ indicates that, for each pair of incident $k_x$ and $\omega$ values, there will be a $k_{1z}$ in medium 1, a $k_{2z}$ in medium 2, and a $k_{3z}$ in medium 3. Given that the dispersion relation specifies each $k_z$ as the square root of a complex entity, there will be a sign ambiguity for each $k_z$ that we will eventually resolve by a proper choice of the corresponding branch-cuts.[9-11]

The branch-points and branch-cuts associated with the $k_z$ components of the various $k$-vectors play a pivotal role in determining the shape of the aforementioned integration contour in the $k_x$-plane, as explained in Sec.6. Also important in deciding the shape of the integration contour in the finite-slab problem are the $k_x$-plane trajectories of the singularities (e.g., poles) of the Fresnel reflection and transmission coefficients; this connection will be elucidated in Sec.7.

Finally, we put the proposed method to use with numerical simulations aimed at computing the reflected and transmitted waves for a finite-duration, finite-width incident wavepacket. A simple model for the incident wavepacket is introduced in Sec.8. (Although, for pedagogical reasons, we use this model of the incident packet in our numerical simulations, we are fully aware of its shortcomings as a realistic model. A more nuanced approach to constructing realistic incident wavepackets is outlined in Appendix A.) Our numerical simulation results are presented in Sec.9, first for a semi-infinite gain medium, and then for a 5.0 micron-thick gainy slab. These simulations are primarily intended to demonstrate the viability of the proposed method of calculation using currently available computational resources. They also reveal the profound differences between the Fresnel reflection and transmission in the presence of gain media versus those involving only passive (i.e., transparent and/or lossy) media. The paper closes with a brief summary of the results and a few concluding remarks in Sec.10.

**2. Spatiotemporal frequencies in the complex plane**. To appreciate a well-known consequence of the causality constraint, consider a linear, shift-invariant system whose output $g(t)$ is related to the input $f(t)$ via a convolution with the system's impulse-response $h(t)$; that is, $g(t) = f(t) * h(t)$. Denoting the frequency by the (real-valued) variable $\omega'$, and the Fourier transforms of our time-dependent functions by $\tilde{f}(\omega')$, $\tilde{g}(\omega')$, and $\tilde{h}(\omega')$, the response of the system to $f(t)$ can be written as follows:[8]

$$g(t) = (2\pi)^{-1}\int_{-\infty}^{\infty} \tilde{f}(\omega')\tilde{h}(\omega')e^{-\mathrm{i}\omega' t}\mathrm{d}\omega'. \tag{1}$$

The above integral, taken from $-\infty$ to $\infty$ along the real axis $\omega'$ of the complex $\omega$-plane, may be regarded as an integral over the lower leg of a contour in the $\omega$-plane that is closed via a large semicircle in the upper half-plane, as shown in Fig.2. If the input $f(t)$ happens to be zero for $t < 0$, then causality dictates that the response $g(t)$ must similarly vanish for $t < 0$. By Cauchy's residue theorem,[9-11] this implies that the integrand in Eq.(1) should be an analytic function of $\omega$ in the upper half of the $\omega$-plane. More specifically, given that $e^{-\mathrm{i}\omega t}$ and $\tilde{f}(\omega)$ are already well-behaved analytic functions — for any reasonable choice of the input $f(t)$ — the implication is that the transfer function $\tilde{h}(\omega)$ should also be analytic in the upper half-plane.

Next, consider a two-dimensional (2D) linear, shift-invariant system whose input $f(x, t)$ is a function of both a spatial coordinate $x$ and the time coordinate $t$. The Fourier expansion of this function may be written as a 2D integral over the real variables $k_x'$ and $\omega'$, as follows:

$$f(x,t) = (2\pi)^{-2}\int_{\omega'=-\infty}^{\infty} \mathrm{d}\omega' e^{-\mathrm{i}\omega' t}\int_{k_x'=-\infty}^{\infty} \tilde{f}(k_x',\omega')e^{\mathrm{i}k_x' x}\mathrm{d}k_x'. \tag{2}$$

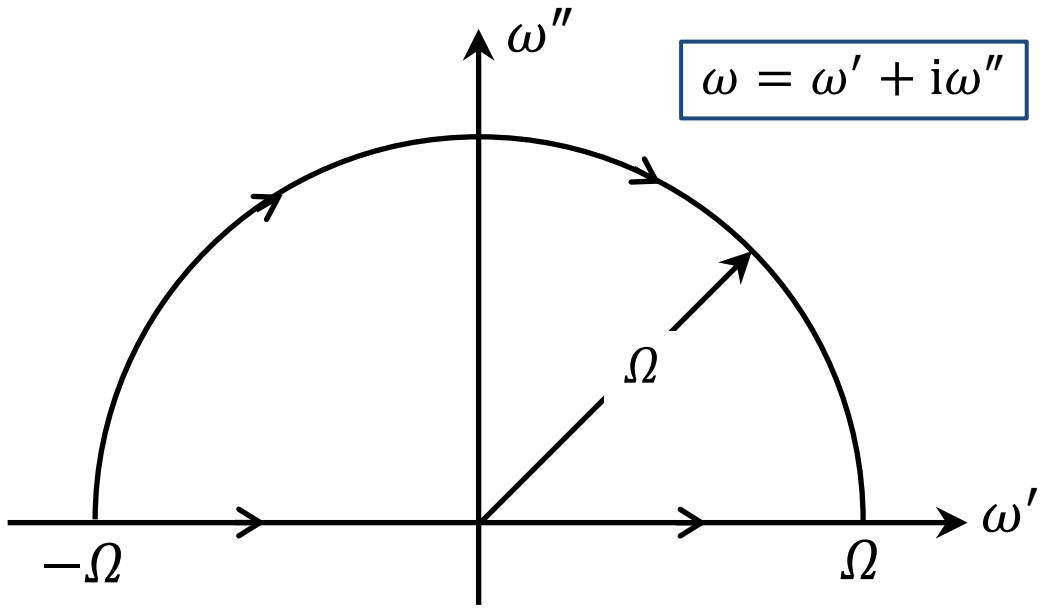

**Fig.2**. According to Cauchy's theorem of complex analysis, the integral from $-\Omega$ to $\Omega$ of an arbitrary function $\phi(\omega)$ on the real-axis $\omega'$ of the $\omega$-plane equals the integral of $\phi(\omega)$ over the depicted semi-circular contour of radius $\Omega$, provided that $\phi(\omega)$ is analytic within the enclosed region between the semi-circle and the straight-line segment from $-\Omega$ to $\Omega$. The preceding statement remains valid in the limit when $\Omega \to \infty$. Thus, if the domain of analyticity of $\phi(\omega)$ is the entire upper-half $\omega$-plane and, furthermore, if $\phi(\omega)$ approaches zero sufficiently rapidly when $|\omega| \to \infty$ in the upper-half plane so that the integral over the semi-circle goes to zero, then Cauchy's theorem guarantees the vanishing of the integral along the real axis, that is, $\int_{-\infty}^{\infty} \phi(\omega')\mathrm{d}\omega' = 0$.

Denoting the system's transfer function in the Fourier domain by $\tilde{h}(k_x', \omega')$, which is also a function of spatial and temporal frequencies $k_x'$ and $\omega'$, the output $g(x,t)$ of the system will be

$$g(x,t) = (2\pi)^{-2} \int_{\omega'=-\infty}^{\infty} \mathrm{d}\omega' e^{-\mathrm{i}\omega' t} \int_{k_x'=-\infty}^{\infty} \tilde{f}(k_x', \omega')\tilde{h}(k_x', \omega')e^{\mathrm{i}k_x' x}\mathrm{d}k_x'. \tag{3}$$

Once again, for causality to hold, the transfer function $\tilde{h}(k_x', \omega)$ should be analytic in the upper-half $\omega$-plane. But now that $\tilde{h}(k_x', \omega)$ is a multivariate function, there are subtleties in the above statement that require careful dissection, because the behavior of $\tilde{h}(k_x', \omega)$ in the $\omega$-plane depends on the value of $k_x'$ at which we evaluate the transfer function. Writing $k_x = k_x' + \mathrm{i}k_x''$ and $\omega = \omega' + \mathrm{i}\omega''$, we must examine $\tilde{h}(k_x, \omega)$ as a function of its four real variables $k_x'$, $k_x''$, $\omega'$, and $\omega''$. In this four-dimensional space, suppose we conduct a search for the set of all points $P_j = (k_x', k_x'', \omega', \omega'')_j$ where $\tilde{h}(k_x, \omega)$ is non-analytic—say, due to the existence of singularities such as poles and/or branch-cuts. If any such points of non-analyticity happen to satisfy $k_x'' = 0$ and $\omega'' \geq 0$ (i.e., at least one $P_j$ lands on the $k_x'$-axis of integration in Eq.(3) *and* in the upper-half of the $\omega$-plane), we may be tempted to declare that $\tilde{h}(k_x', \omega)$ is not analytic in the upper-half $\omega$-plane and that, therefore, the causality of the system cannot be assured; see Fig.3.

That conclusion, however, would be premature; we can rescue causality, because we have the freedom to switch to an alternative integration path in the $k_x$-plane—to be justified shortly. By moving the integration path away from the real $k_x'$-axis and onto a contour that bypasses all $k_x$-plane singularities $(k_x', k_x'')_j$ associated with $\omega_j'' \geq 0$, one can restore analyticity to the upper-half $\omega$-plane, thereby affirming the causality of the system; a typical deformed integration path in the $k_x$-plane is shown in Fig.3.

In summary, the recipe for treating such problems is:

i) Find all points $P_j = (k_x', k_x'', \omega', \omega'')_j$ where $\omega_j'' \geq 0$, and the resultant $\tilde{h}(k_x, \omega)$ is non-analytic (say, due to the existence of a singularity such as a pole or a branch-cut).

ii) Identify the projections $(k_x', k_x'')_j$ into the $k_x$-plane of all the singular points $P_j$ with $\omega_j'' \geq 0$.

iii) Choose an integration contour in the $k_x$-plane that bypasses all such points of non-analyticity.

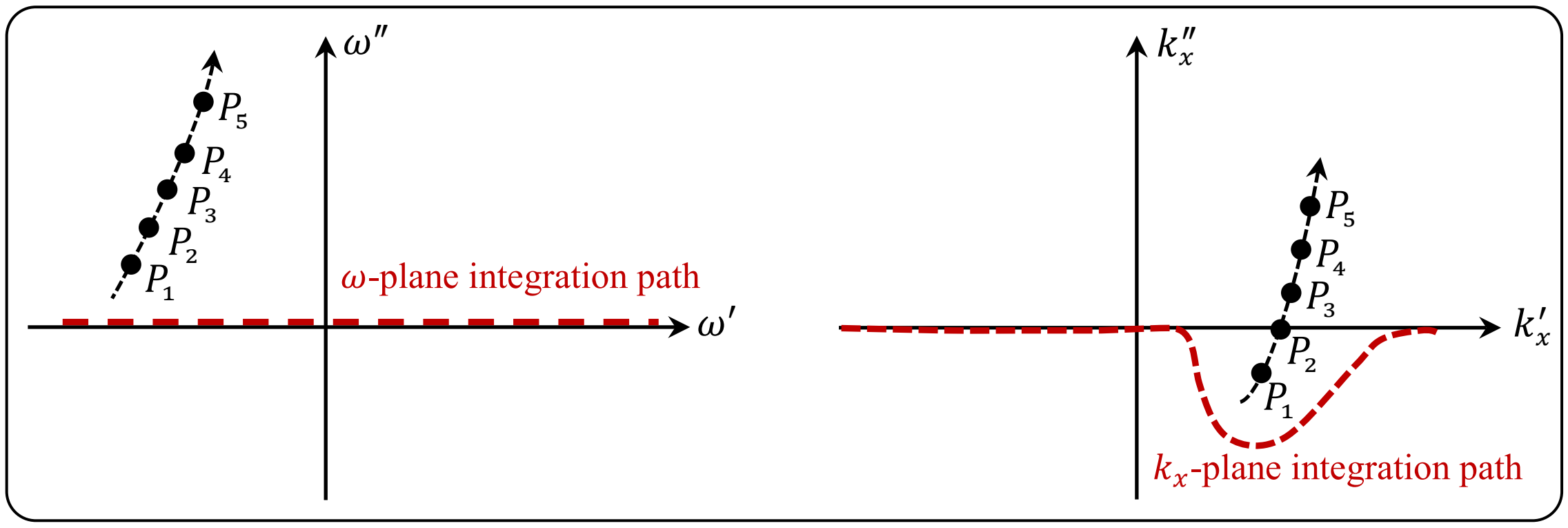

**Fig.3**. The transfer function $\tilde{h}(k_x, \omega)$ is a function of the four variables $k_x', k_x'', \omega', \omega''$. Since complex-plane integration is carried out on specific contours in the planes of individual variables $k_x = k_x' + \mathrm{i}k_x''$ and $\omega = \omega' + \mathrm{i}\omega''$, it is natural to consider the projections of each singular point $P_j$ into the $\omega$-plane and the $k_x$-plane. In this example, the singularities begin at $P_1$, then follow a continuous trajectory (along the dashed black lines) in both the $\omega$-plane and the $k_x$-plane. There exists a continuum of such singularities, of which five points along the trajectory have been highlighted. The singularities whose projections lie in the upper half of the $\omega$-plane are rendered moot by choosing an integration contour in the $k_x$-plane that bypasses those singularities. In this way, the transfer function remains analytic in the upper-half $\omega$-plane, which is needed to ensure the overall causality of the response of the system.

The question that must be asked at this point is: Why do we have the freedom to deform the integration path in the $k_x$-plane when evaluating Eq.(3) for the system output $g(x, t)$? Shouldn't the Fourier integral be taken along the real $k_x'$-axis? The answer is that one must use the same integration path to evaluate $g(x, t)$ in Eq.(3) as that used to evaluate the input $f(x, t)$ in Eq.(2). So now we must justify the freedom to deform the $k_x$-plane integration contour in Eq.(2). To this end, observe that the 2D Fourier transform of the input function is given by

$$\tilde{f}(k_x, \omega) = \int_{t=-\infty}^{\infty} \mathrm{d}t e^{\mathrm{i}\omega t} \int_{x=-\infty}^{\infty} f(x, t) e^{-\mathrm{i}k_x x} \mathrm{d}x. \quad (4)$$

Here, there is no difficulty in evaluating the integrals along the real $x$ and real $t$ axes provided that $f(x, t)$ is sufficiently well-behaved. Moreover, there exist circumstances where the domain of $\tilde{f}(k_x, \omega)$ can be extended to encompass complex values of $k_x$ and $\omega$, as, for instance, when $f(x, t)$ has finite duration in time and finite width along the $x$-axis. Now, the standard choice of contours to evaluate the inverse Fourier integrals in Eq.(2) is along the real $k_x'$ and real $\omega'$ axes. Nevertheless, if $\tilde{f}(k_x, \omega)$ happens to be analytic within the area enclosed between the $k_x'$-axis and a deformed contour $C$ in the $k_x$-plane, nothing would prevent us from switching the $k_x$-plane integration path from the real $k_x'$-axis to the contour $C$. Thus, Eq.(2) may be written as

$$f(x, t) = (2\pi)^{-2} \int_{\omega'=-\infty}^{\infty} \mathrm{d}\omega' e^{-\mathrm{i}\omega' t} \int_C \tilde{f}(k_x, \omega') e^{\mathrm{i}k_x x} \mathrm{d}k_x, \quad (5)$$

where the integration path along the $k_x'$-axis has been replaced with an appropriate contour $C$ within the $k_x$-plane. Physically, this means that the input function $f(x, t)$ does *not* have a unique expansion into plane-waves of the general form $e^{\mathrm{i}(k_x x - \omega' t)}$; there exist infinitely many such superpositions that are equally valid, one for each choice of the contour $C$. Later, when we need to calculate the system output $g(x, t)$ via Eq.(3), we capitalize on that freedom to choose a particular superposition (i.e., a particular contour $C$ in the $k_x$-plane) that preserves the analyticity of the inverse Fourier integrand within the upper-half $\omega$-plane. Consequently, this choice of the integration contour guarantees the causality of the output $g(x, t)$ of the system in response to the input $f(x, t)$.

We have presented the above method in a way that suggests the same $k_x$-plane integration contour $C$ must be used for each and every value of $\omega'$ in the $\omega$-integral. As helpful as this may be as a way of thinking about the procedure for didactic purposes, in practice, it is not an actual restriction. When evaluating the inverse Fourier integral numerically, one may pick different $k_x$-plane integration contours for each value of $\omega'$, if that happens to be more convenient.

Finally, it must be pointed out that, in the aforementioned step iii, one may *not* be able to choose an integration path that bypasses all the singular points in the $k_x$-plane associated with the upper-half $\omega$-plane. When this happens, the method fails and the system is said to exhibit an absolute instability — as opposed to a convective instability.[12-15]

This section has provided a bird's eye view of the fundamental mathematical procedure used in the remainder of the paper. In what follows, we apply this method to the Fresnel problem to compute the reflected and transmitted wavepackets in the presence of a gain medium. Many important details that have been mentioned briefly here will be brought up again and expanded upon as our solution to the Fresnel problem is elaborated.

**3. Statement of the problem**. Having outlined a possible approach to ensuring the causality of a system whose physical evolution unfolds in space as well as in time, we now cast the Fresnel problem in this framework. The two systems of interest in this paper are schematically shown in Fig.1. The incident wavepacket is uniform along the $y$-axis, thus rendering our analysis independent of the $y$ coordinate. The fixed observation point $(x_0, z_0)$ may be located in medium 1 (where $z_0 \le 0$), or medium 2 (where $z_0 \ge 0$ for the single-surface problem, and $0 \le z_0 \le d$ for the finite-slab problem), or, in the case of the finite slab, in medium 3 (where $z_0 \ge d$). All three media are linear, isotropic, homogeneous, and non-magnetic, specified by their dielectric permittivity $\varepsilon(\omega)$ and magnetic permeability $\mu(\omega) = 1$. Thus, the optical properties of the $j^{\text{th}}$ medium are encapsulated in its dielectric function $\varepsilon_j(\omega)$, which is related to the refractive index via the standard identity $n_j(\omega) = \sqrt{\varepsilon_j(\omega)}$.[1] The goal is to compute the reflected and/or transmitted $E$-field at the observation point as a function of time, namely, $\boldsymbol{E}(x_0, z_0, t)$.

Given that we will invoke the causality constraint to settle the choice of the wave-vector, this problem cannot be solved by considering only idealized monochromatic plane-waves, which are spatially unbounded and have no definite starting point in time. Instead, we consider a realistic wavepacket arriving at the interfacial $xy$-plane located at $z = 0$ with a finite footprint (i.e., illuminated area) along the $x$-axis and a definite starting point in time, which, without loss of generality, we may assume to be at $t = 0$. Thus, the EM fields of the incident packet arriving at the $xy$-plane at $z = 0$ are precisely zero for $t < 0$ and, for $t \ge 0$, they vanish outside a finite interval $[x_{\min}, x_{\max}]$ along the $x$-axis. (This is all the information needed here to formulate the theoretical approach; the specific form of the incident packet used in our numerical simulations is described in Sec.8.)

Under such circumstances, the incident packet having the electric field $\boldsymbol{E}^{(\text{inc})}(x, t)$ at $z = 0$, can be described as a superposition of plane-waves whose $E$-fields are $\widetilde{\boldsymbol{E}}^{(\text{inc})}(k_x, \omega) \exp[\mathrm{i}(k_x x + k_{1z} z - \omega t)]$, where the temporal frequency $\omega$ and the spatial frequency $k_x$ assume all possible values from $-\infty$ to $\infty$. Here, $\widetilde{\boldsymbol{E}}^{(\text{inc})}(k_x, \omega)$ is the 2D Fourier transform of $\boldsymbol{E}^{(\text{inc})}(x, z = 0, t)$. We confine our attention to the case of transverse electric or TE (also known as $s$-polarized) waves, so that the $E$-fields everywhere can be written as $E_y(x, z, t)\hat{\boldsymbol{y}}$, although the employed methods are quite general and can readily be applied to incident beams having other states of polarization.

The $z$-component of the $k$-vector associated with $(k_x, \omega)$ in medium $j$ is given by the dispersion relation — itself a consequence of Maxwell's equations — as follows:[1-3]

$$k_{jz} = \pm(\omega/c)\sqrt{\varepsilon_j(\omega) - (ck_x/\omega)^2}. \tag{6}$$

Note that the plus or minus sign of $k_{jz}$ in Eq.(6) remains to be specified. The incident packet arrives at either the single-surface or the finite-slab configuration depicted in Fig.1. We start by writing down the equations governing the system's response, and proceed to analyze each piece of these equations in detail. In the single-surface problem, for instance, the reflected and transmitted $E$-fields at the observation point $(x_0, z_0)$ are obtained via the following inverse Fourier transform relations:

$$E_y^{(\text{ref})}(x_0, z_0, t) = (2\pi)^{-2} \int_{\omega'=-\infty}^{\infty} \mathrm{d}\omega' e^{-\mathrm{i}\omega' t} \int_C \tilde{E}_y^{(\text{inc})}(k_x, \omega')\, \rho(k_x, \omega') e^{\mathrm{i}(k_x x_0 - k_{1z} z_0)} \mathrm{d}k_x, \quad (7)$$

$$E_y^{(\text{trans})}(x_0, z_0, t) = (2\pi)^{-2} \int_{\omega'=-\infty}^{\infty} \mathrm{d}\omega' e^{-\mathrm{i}\omega' t} \int_C \tilde{E}_y^{(\text{inc})}(k_x, \omega')\, \tau(k_x, \omega') e^{\mathrm{i}(k_x x_0 + k_{2z} z_0)} \mathrm{d}k_x. \quad (8)$$

In connection with the discussion in Sec.2, the transfer functions here are $\rho(k_x, \omega') e^{-\mathrm{i}k_{1z} z_0}$ and $\tau(k_x, \omega') e^{\mathrm{i}k_{2z} z_0}$. In these equations, $k_{1z}$ and $k_{2z}$ are the $z$-components of the $k$-vectors within the incidence and transmittance media, which — aside from a sign ambiguity which we set out to resolve — are determined by the dispersion relation in Eq.(6).[1-3] The Fresnel reflection and transmission coefficients $\rho$ and $\tau$ appearing in Eqs.(7) and (8) are themselves functions of $k_{1z}$ and $k_{2z}$, which are, in turn, related to $k_x$ and $\omega$; these relations will be discussed in some detail in the following sections. The guiding principle is to understand the behavior of the inverse Fourier integrands of Eqs.(7) and (8) in the complex $k_x$ and $\omega$ planes, and to choose a $k_x$-plane contour $C$ that renders the integrands analytic in the upper half $\omega$-plane. As it turns out, the single-surface problem depicted in Fig.1(a) provides an instructive example for how to deal with the various $k_z$ branch-cuts as the cause of non-analyticity of the integrand, whereas the finite-slab problem of Fig.1(b) exemplifies the method of dealing with other types of singularity (e.g., poles) of the corresponding integrands.

**4. Lorentz oscillators**. In general, each of the three media, modelled as a collection of $K$ Lorentz oscillators, has its own dielectric permittivity $\varepsilon(\omega)$, as follows:[1-3]

$$\varepsilon(\omega) = 1 + \sum_{\kappa=1}^{K} f_\kappa \omega_{p\kappa}^2 / (\omega_{r\kappa}^2 - \omega^2 - \mathrm{i}\gamma_\kappa \omega). \quad (9)$$

The above equation contains the standard plasma frequency $\omega_p$, resonance frequency $\omega_r$, damping coefficient $\gamma$, and oscillator strength $f$ for each oscillator. If the resonance line-widths are sufficiently narrow (i.e., small $\gamma_\kappa$) and the resonance frequencies $\omega_{r1}, \omega_{r2}, \cdots, \omega_{rK}$ are sufficiently far apart, the various oscillators act more or less independently of each other. Each oscillator then dominates a range of frequencies centered at $\omega_{r\kappa}$. Each medium has its own set of $4K$ parameters $(f_\kappa, \omega_{p\kappa}, \omega_{r\kappa}, \gamma_\kappa)$, with $\kappa$ ranging from 1 to $K$. In the case of the passive media (1 and 3), all oscillator strengths $f_\kappa$ are $+1$, whereas the active medium (i.e., gain medium 2) has at least one oscillator whose strength $f_\kappa$ equals $-1$.[†]

The general aspects of the problem can be studied for single-oscillator media (i.e., $K = 1$). Multi-oscillator media are not expected to introduce conceptual or mathematical difficulties — at least in cases where the various resonance frequencies of each medium are sufficiently far apart from one another — beyond those already encountered in the case of single-oscillator media. Therefore, we will use the dielectric functions $\varepsilon_1(\omega)$ and $\varepsilon_2(\omega)$ of media 1 and 2 given by the single-oscillator Lorentz model,[2,3] as follows:

[†]For non-magnetic media, where the relative permeability $\mu(\omega)$ is 1, the refractive index $n(\omega)$, the dielectric susceptibility $\chi(\omega)$, and the relative permittivity $\varepsilon(\omega)$ are related via the standard identity $n(\omega) = \sqrt{\mu(\omega)\varepsilon(\omega)} = \sqrt{1 + \chi(\omega)}$.

$$\varepsilon_1(\omega) = 1 + \frac{\omega_{p1}^2}{\omega_{r1}^2 - \omega^2 - \mathrm{i}\gamma_1\omega}; \qquad \varepsilon_2(\omega) = 1 - \frac{\omega_{p2}^2}{\omega_{r2}^2 - \omega^2 - \mathrm{i}\gamma_2\omega}. \tag{10}$$

Note that the passive medium 1 is distinguished from the active (gain) medium 2 by the plus sign versus the minus sign appearing immediately after 1 on the right-hand side.[16] The dielectric function $\varepsilon_3(\omega)$ of medium 3 is similar to $\varepsilon_1(\omega)$, albeit with its own parameter set $(\omega_{p3}, \omega_{r3}, \gamma_3)$.

Although the Lorentz oscillators obey the Kramers-Kronig relations, thereby guaranteeing the causal response of the individual dipoles of each medium, the technique described in Sec.2 is still needed to ensure that the system as a whole complies with the causality constraint.

**5. Reflection and transmission coefficients**. For an $s$-polarized incident wavepacket (i.e., one whose $E$-field is aligned with the $y$-axis, also known as a transverse electric or TE wave), the Fresnel reflection and transmission coefficients in the single-surface problem of Fig.1(a) are[1-3]

$$\rho(k_x, \omega) = (k_{1z} - k_{2z})/(k_{1z} + k_{2z}), \tag{11}$$

$$\tau(k_x, \omega) = 2k_{1z}/(k_{1z} + k_{2z}). \tag{12}$$

In the finite-slab problem of Fig.1(b), the reflection coefficient at the slab's front facet $(z = 0)$ and transmission coefficient at its rear facet $(z = d)$ are given by[1-3]

$$\rho(k_x, \omega) = \frac{\rho_{12} + \rho_{23}\exp(2\mathrm{i}k_{2z}d)}{1 - \nu}, \tag{13}$$

$$\tau(k_x, \omega) = \frac{(1 + \rho_{12})(1 + \rho_{23})\exp(\mathrm{i}k_{2z}d)}{1 - \nu}. \tag{14}$$

The transmission coefficient yielding the $E$-field inside the gain layer (medium 2) at $z = z_0$ is

$$\tau(k_x, \omega) = \frac{(1+\rho_{12})\{\exp(\mathrm{i}k_{2z}z_0) + \rho_{23}\exp[\mathrm{i}k_{2z}(2d - z_0)]\}}{1 - \nu}. \tag{15}$$

In Eqs.(13)-(15), the roundtrip coefficient $\nu$, itself a function of $k_x$ and $\omega$, is given by

$$\nu = \rho_{21}\rho_{23}\exp(2\mathrm{i}k_{2z}d). \tag{16}$$

For an $s$-polarized incident wave, the reflection coefficients $\rho_{12}$, $\rho_{21}$, and $\rho_{23}$, are known to be[1-3]

$$\rho_{12} = -\rho_{21} = (k_{1z} - k_{2z})/(k_{1z} + k_{2z}), \tag{17}$$

$$\rho_{23} = (k_{2z} - k_{3z})/(k_{2z} + k_{3z}). \tag{18}$$

The $z$-components of the $k$-vectors within media 1, 2, and 3, denoted by $k_{1z}$, $k_{2z}$, $k_{3z}$ and given by Eq.(6), will be described in the following section. For now, it is important to recognize that the various $k_z$s are functions of both $\omega$ and $k_x$, where $\omega$ and $k_x$ are generally complex-valued. Each $k_z$, being the square root of a complex entity, requires a choice of plus or minus sign, with only one of the two signs being acceptable at any given point $(k_x, \omega)$.[‡] One must ensure that $k_z(-k_x^*, -\omega^*) = -k_z^*(k_x, \omega)$, in order to guarantee the Hermitian symmetry relations $\rho(-k_x^*, -\omega^*) = \rho^*(k_x, \omega)$ and $\tau(-k_x^*, -\omega^*) = \tau^*(k_x, \omega)$, which are essential if the reflected and transmitted fields at all observation points are to be real-valued.

[‡]The only exception to this rule is the sign of $k_{2z}$ in the case of the finite-slab of Fig.1(b), where Eqs.(13)-(18) yield the same values for $\rho(k_x, \omega)$ and $\tau(k_x, \omega)$ irrespective of the chosen sign for $k_{2z}$. Consequently, the choice of branch-cuts for $k_{2z}$ in the finite-slab problem is of no significance. In all other cases, one must carefully choose the branch-cuts for $k_{1z}$, $k_{2z}$, and $k_{3z}$, to ensure that each acquires its correct sign.

In connection with the reflected $E$-field computed in accordance with Eq.(7), the Fresnel reflection coefficient $\rho$ is given by Eq.(11) in the case of a semi-infinite gain medium, and by Eq.(13) in the case of a finite-thickness slab. Similarly, the transmitted $E$-field inside the semi-infinite gain medium of Fig.1(a) is obtained from Eq.(8) using the Fresnel transmission coefficient $\tau$ given by Eq.(12). In the case of the finite-thickness slab of Fig.1(b), the transmitted $E$-field inside medium 3 is obtained using the Fresnel coefficient $\tau$ of Eq.(14), provided that $k_{3z}(z_0 - d)$ is substituted for $k_{2z}z_0$ within the exponential factor in Eq.(8). As for the $E$-field inside the slab itself, one must use Eq.(8) in conjunction with $\tau(k_x, \omega')$ of Eq.(15) and, given that Eq.(15) already incorporates the relevant propagation phase-factor within the slab, the term $k_{2z}z_0$ should be removed from the exponential factor in Eq.(8).

Inside the complex $\omega$-plane, the region of interest will be the upper half-plane, although certain symmetries allow us to focus our attention exclusively on the first quadrant ($Q_1$), where $\omega = \omega' + \mathrm{i}\omega''$ has $\omega' \geq 0$ and $\omega'' \geq 0$. For all points in the second quadrant ($Q_2$), we have $\varepsilon(-\omega^*) = \varepsilon^*(\omega)$, a direct consequence of the Lorentz oscillator model of Eq.(9). For these $Q_2$ points of the $\omega$-plane, if we switch $k_x$ to $-k_x^*$, and ensure that the corresponding $k_z$ also switches to $-k_z^*$, then the $E$-field at the observation point $(x_0, z_0)$ acquires the conjugate of its value at the corresponding frequency in $Q_1$ of the $\omega$-plane. Therefore, there will be no need to keep track of the points in $Q_2$ of the $\omega$-plane. (The symmetry between $Q_1$ and $Q_2$ of the $\omega$-plane also implies that the contribution to the $E$-field at $(x_0, z_0)$ by frequencies $\omega$ that reside on the positive imaginary axis $\omega''$ must be real-valued.)

The bulk of our computational effort revolves around identifying and then eliminating the singularities of the inverse Fourier integrands of Eqs.(7) and (8) from $Q_1$ of the $\omega$-plane. As pointed out in Sec.2, removing all these singularities is necessary to ensure that causality is satisfied. It is worth emphasizing here that the final step in the calculation of $E_y(x_0, z_0, t)$ is an inverse Fourier transformation over the real $\omega'$-axis, as seen in Eqs.(7) and (8). Considering that the contribution to the inverse Fourier integral of the negative half of this axis equals the conjugate of that from the positive half, one can simplify the calculation by taking the real part of the integral computed only for $\omega' = 0$ to $\infty$.

Having identified all the terms of the integrands of Eqs.(7) and (8), we now examine these integrands for their regions of non-analyticity due to the existence of branch-cuts, and then poles.

**6. Branch-points and branch-cuts**. As discussed in Sec.2, it is necessary to identify all the non-analytic points (i.e., poles and branch-cuts) of the integrands in Eqs.(7) and (8) that satisfy $\omega'' \geq 0$. Focusing on the branch-points in the present section, we note that the wave-vector components $k_{1z}$, $k_{2z}$, and $k_{3z}$ appear throughout the inverse Fourier integrands, both explicitly (in the exponential terms) and implicitly (in the Fresnel coefficients). Since the function $k_{jz}(k_x, \omega)$ involves a square root, a branch-cut is needed to uniquely evaluate this square root. To examine the branch-points and branch-cuts in the $k_x$-plane, we decompose $k_{jz}$ into the following product of two square roots:

$$k_{jz} = \sqrt{(\omega/c)^2 \varepsilon_j(\omega) - k_x^2} = -\mathrm{i}[k_x - \omega n_j(\omega)/c]^{1/2}[k_x + \omega n_j(\omega)/c]^{1/2}. \quad (19)$$

Recall that, for a given $\omega' \geq 0$, $n_1(\omega')$ and $n_3(\omega')$ will each have a value in $Q_1$ and a symmetrically located value in $Q_3$ of the complex plane — a simple consequence of the fact that $n_1$ and $n_3$ are the square roots of $\varepsilon_1$ and $\varepsilon_3$, both of which are $Q_1$ complex numbers. In the case of $n_2(\omega')$, if the dominant oscillator in the vicinity of $\omega'$ happens to be gainy (i.e., $f_k < 0$), then the values of $n_2(\omega')$ will be in $Q_2$ and $Q_4$; otherwise, $n_2(\omega')$ will be in $Q_1$ and $Q_3$.

We start by choosing a fixed point on the $\omega$-plane integration path with $\omega' \geq 0$ and $\omega'' = 0$. Each of the two radicals on the right-hand side of Eq.(19) will then have a branch-point[9,10] within the $k_x$-plane at $k_x = \pm\omega' n_j(\omega')/c$. For $k_{1z}$ and $k_{3z}$, these branch-points always end up in $Q_1$ and $Q_3$ of the $k_x$-plane. For $k_{2z}$, the branch-points will be in $Q_2$ and $Q_4$ of the $k_x$-plane if $\omega'$ happens to be near the resonance frequency of a gainy oscillator; otherwise, they will land in $Q_1$ and $Q_3$. Having found the location of the branch-points for $\omega'' = 0$, we now trace the trajectories of these branch-points as $\omega''$ rises from 0 to $+\infty$, the reason being the need to identify all possible non-analytic points in the upper-half $\omega$-plane — not just those that fall on the $\omega$-plane integration path. The $k_x$-plane trajectories of the branch-points of $k_{1z}$, $k_{2z}$, and $k_{3z}$ for a fixed value of $\omega' > 0$ are shown schematically in Fig.4 (and computed precisely for a specific example in Sec.9). The branch-point trajectories in $Q_1$ and $Q_4$ move upward, whereas the corresponding trajectories in $Q_2$ and $Q_3$ move downward (due to the inherent odd symmetry).

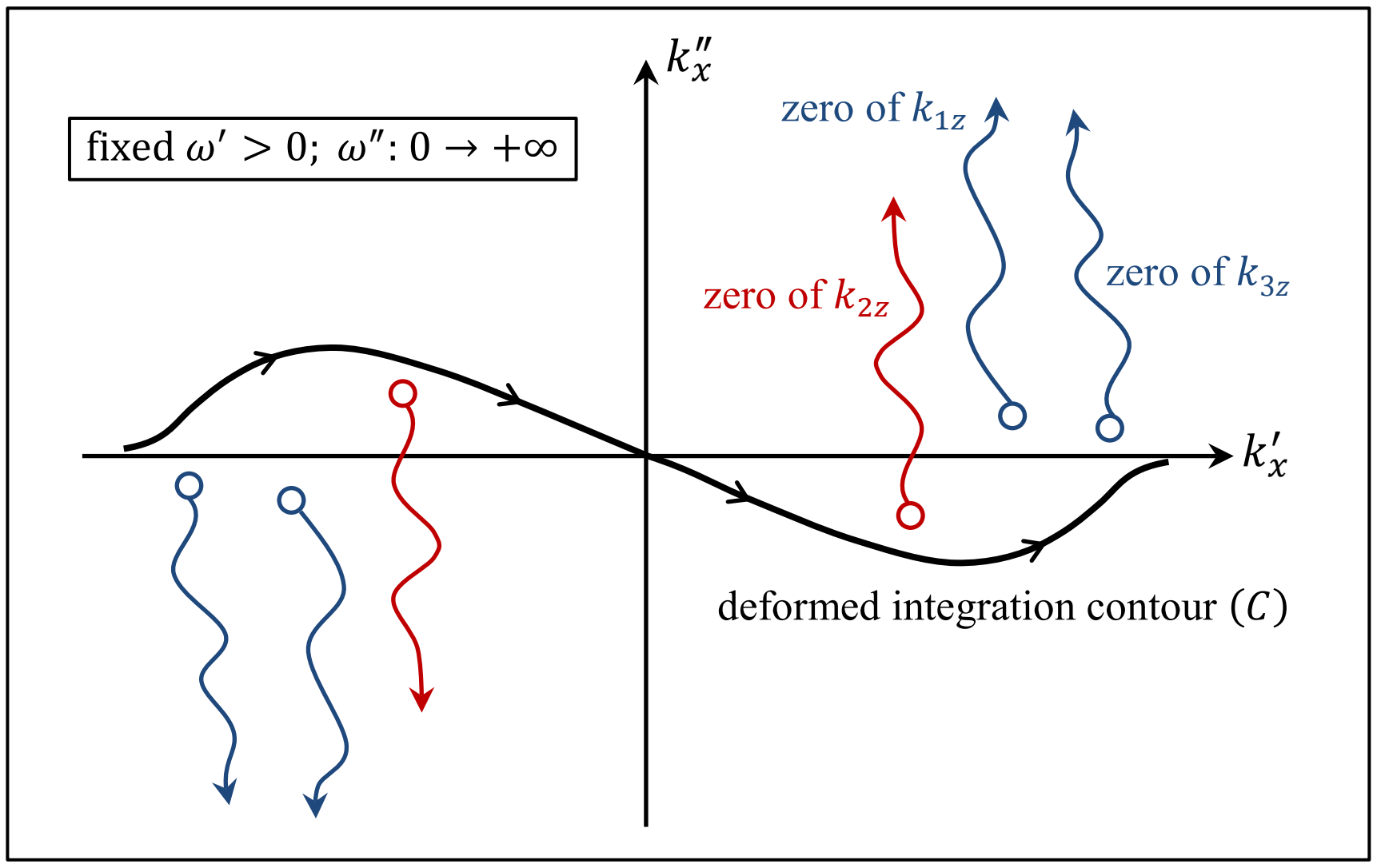

**Fig. 4**. Trajectories of the zeros of $k_{1z}$, $k_{2z}$, and $k_{3z}$ in the complex $k_x$-plane, when $\omega'$ is fixed at a positive value while $\omega''$ rises from 0 to $+\infty$. In compliance with the requirement to bypass all the singular points, the chosen integration contour $C$ avoids crossing the branch-point trajectories for all values of $\omega' \geq 0$.

The branch-cuts should be drawn as straight vertical lines, originating from each branch-point and extending to infinity — although, to reduce clutter, some may find it convenient to designate the branch-point trajectories themselves as branch-cuts.[11] Note that the branch-point trajectories for $k_{1z}$ and $k_{3z}$ never cross the $k_x'$-axis, whereas the trajectories for $k_{2z}$ that start in $Q_2$ and $Q_4$ (i.e., those corresponding to gainy oscillators) do cross the $k_x'$-axis. These branch-point trajectories for $k_{2z}$ necessitate a deformation of the integration contour away from the $k_x'$-axis in the $k_x$-plane (i.e., inverse Fourier transformation with respect to $k_x$), lest the integration path crosses a branch-cut, which would result in a discontinuity of $k_{2z}$ that heralds a violation of causality by rendering the integrand in the upper-half $\omega$-plane non-analytic.

One such deformed contour, though by no means a unique choice, appears in Fig.4. Note that one could choose a different $k_x$-contour for each $\omega'$ along the $\omega$-contour. Alternatively, one could draw the branch-point trajectories of Fig.4 for *all* values of $\omega' \geq 0$, then choose a single $k_x$-contour that bypasses all of them at once. We choose the latter method in our numerical simulations, though, in the end, either method would give the same result. By avoiding the branch-point trajectories in the $k_x$-plane, our chosen integration path ensures the analyticity of the wave-vector components $k_{jz}(k_x, \omega)$ in all three media.

Lastly, we must specify the range of the phase angles $\varphi$ at the branch-cuts associated with the individual terms whose square roots appear on the right-hand side of Eq.(19). A typical example explaining the use of one such pair of branch-cuts for $k_{2z}$ is shown in Fig.5. For the branch-cut depicted on the left-hand side, $\varphi$ ranges from $-90°$ to $270°$, whereas that on the right confines the corresponding $\varphi$ to the interval from $90°$ to $450°$. To calculate $k_{2z}$ at a given point $k_x$ on the contour $C$, draw the two green arrows from each branch-point to $k_x$; these represent the two terms whose square roots appear on the right-hand side of Eq. (19). With the phases of these two numbers determined by the respective branch-cuts, each square root becomes a one-to-one function that takes the complex number $|a|e^{\mathrm{i}\varphi}$ to $|a|^{½}e^{\mathrm{i}\varphi/2}$. Thus, by uniquely specifying $k_{2z}$ everywhere along the $k_x$-contour, the branch-cuts eliminate the sign ambiguity in the dispersion relation. Additionally, since the contour $C$ is chosen to bypass all branch-cuts, $k_{2z}$ varies continuously with $k_x$ and $\omega$, becoming an analytic function in the upper-half of the $\omega$-plane.

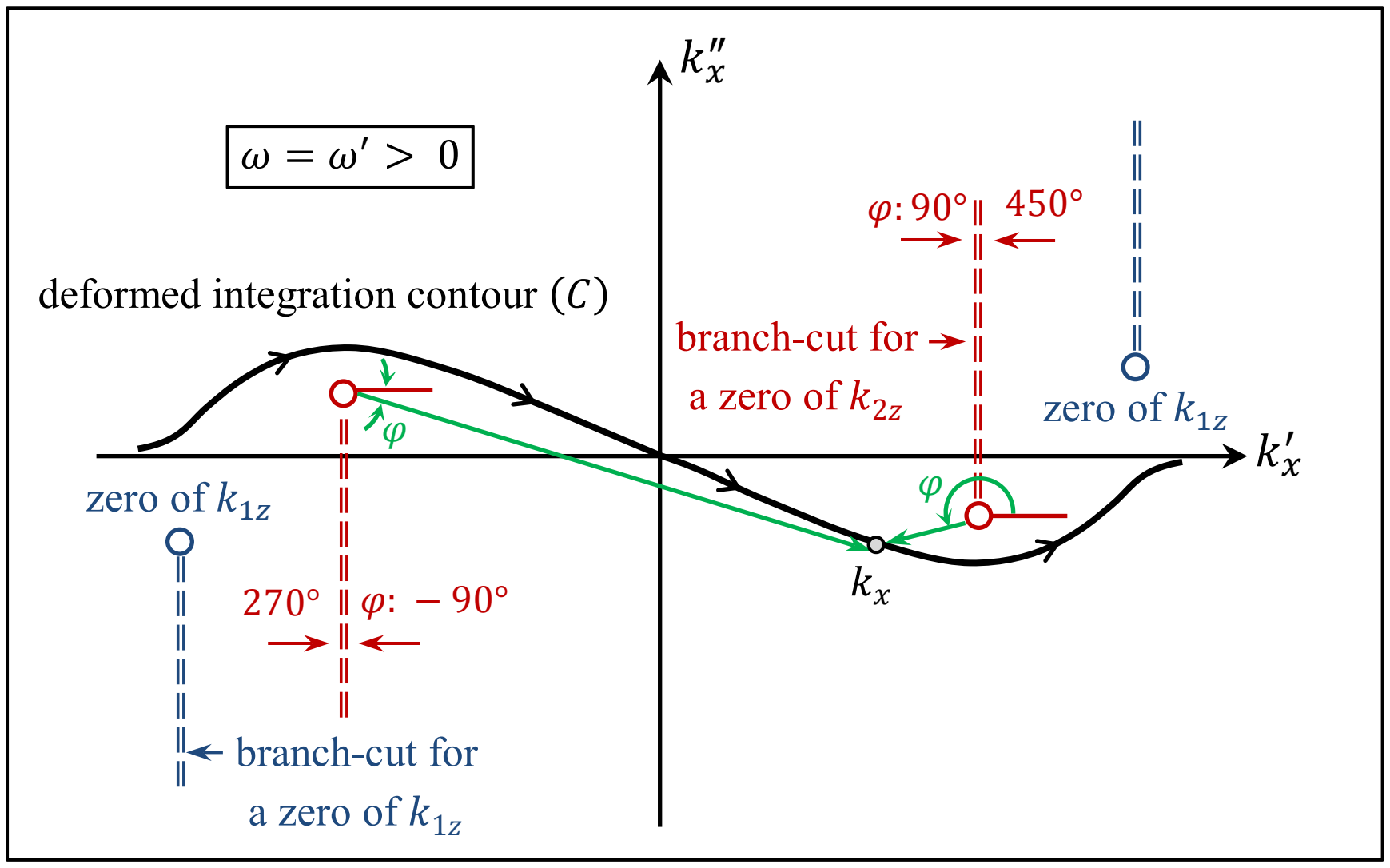

**Fig.5**. Branch-cuts for $k_{1z}$ and $k_{2z}$ in the $k_x$-plane corresponding to a fixed, positive, real-valued frequency $\omega$. The branch-cuts are chosen to ensure that they do not cross the deformed integration contour $C$. For a specific (real and positive) $\omega$ and a specific point $k_x$ on the integration contour, the solid green arrows represent the two terms whose square roots appear in Eq.(19) in the expression of $k_{2z}$. The chosen range for the pair of angles $\varphi$ guarantees that $k_{1z}$, $k_{2z}$, $k_{3z}$ approach $\omega/c$ in the limit when $\omega'' \to +\infty$.

The above choices for the phase angles $\varphi$ were made to ensure that $k_{2z} \to \omega/c$ in the limit when $\omega'' \to +\infty$. The necessity for all three $k_z$s to approach $\omega/c$ when $\omega'' \to +\infty$ can be appreciated by examining the exponential factors in the integrands of Eqs.(7) and (8) in conjunction with the semi-circular integration path in the upper-half $\omega$-plane of Fig.2, on which the integrands are required to vanish for $t < |z_0|/c$.

Interestingly, the branch-cuts for $k_{2z}$ play an important role in determining the integration contour $C$ only in the single-surface problem depicted in Fig.1(a). For a gainy slab of finite thickness $d$, such as that of Fig.1(b), the Fresnel reflection and transmission coefficients of the slab are insensitive to the choice of sign for $k_{2z}$ (i.e., the coefficients remain the same under the transformation $k_{2z} \to -k_{2z}$); consequently, the branch-cuts of $k_{2z}$ play no role whatsoever in determining the $k_x$-plane integration contour $C$ for the finite-slab problem. In this case, the choice of integration contour is dictated by other singularities of the Fresnel reflection and transmission coefficients whose $\omega$-values happen to be in $Q_1$ of the $\omega$-plane. These singularities are discussed in the next section.

**7. Singularities of the Fresnel reflection and transmission coefficients**. In the case of a gainy slab of finite-thickness $d$, when the roundtrip coefficient $\nu$ defined in Eq.(16), itself a function of $k_x$ and $\omega$, equals 1.0, the Fresnel reflection and transmission coefficients ($\rho$ and $\tau$) diverge to infinity, causing the inverse Fourier integrands in Eqs.(7) and (8) to exhibit a singularity at the corresponding point $(k_x, \omega)$. Considering that $\nu$ is a function of $k_x^2$, the $k_x$-values of the singularities always appear in symmetric pairs (i.e., $\pm k_x$). In general, a large number of such singularities exist, and one must conduct an exhaustive numerical search to identify all such singular points $(\pm k_x, \omega)$ for all $Q_1$ points of the $\omega$-plane.

In the single-surface problem of Fig.1(a), the singularities of $\rho(k_x, \omega)$ and $\tau(k_x, \omega)$ will be the zeros of $k_{1z} + k_{2z}$; see Eqs.(11) and (12). If both media 1 and 2 happen to have only a single Lorentz oscillator, it can be shown that $\rho$ and $\tau$ will have no singularities in the upper-half $\omega$-plane. In the case of multi-oscillator media, however, a numerical search must be conducted to identify the singularities $(\pm k_x, \omega)$ associated with all points in $Q_1$ of the $\omega$-plane. The $k_x$-plane integration contour $C$ must then bypass the $\pm k_x$ singular points residing in $Q_2$ and $Q_4$ of the $k_x$-plane, in addition to avoiding all the branch-point trajectories of $k_{2z}$, as described previously.

Due to symmetries that allow the $\omega$-integral to be evaluated along the positive semi-axis $\omega' > 0$, we only need to find the singularities that fall within $Q_1$ of the $\omega$-plane. Most of the $\pm k_x$ pairs associated with singular points $\omega$ in $Q_1$ of the $\omega$-plane turn up in $Q_1$ and $Q_3$ of the $k_x$-plane. However, for a weakly-amplifying medium, a limited number of such singular $\pm k_x$ pairs appear in $Q_2$ and $Q_4$ of the $k_x$-plane, not too far away from the real $k_x'$-axis. By picking a deformed contour $C$ in the $k_x$-plane in such a way as to avoid all such singularities — say, by staying above all singular points in $Q_2$ and below all singular points in $Q_4$, in the same manner as we bypassed the branch-point trajectories in Fig.4 — one can guarantee the absence of singularities in $Q_1$ of the $\omega$-plane. In this way, upon evaluating the inverse Fourier integrals over the $k_x$ variable in Eqs.(7) and (8) along a properly deformed contour $C$, the resulting function of $\omega$ ends up being analytic throughout the entire $Q_1$ of the $\omega$-plane. (The analyticity of this function in $Q_2$ automatically follows from its mirror symmetry with respect to the imaginary $\omega''$-axis.)

The fact that, for any given point in the upper-half $\omega$-plane, the corresponding singularities in the $k_x$-plane appear as $\pm k_x$ pairs, accounts for the odd symmetry of the contour $C$ with respect to the $k_x''$-axis, as depicted in Figs.4 and 5. Of course, the contour need not obey the same odd symmetry as the singularities, but is only required to properly bypass these non-analytic points.

We have previously pointed out that the symmetry between $Q_1$ and $Q_2$ of the $\omega$-plane obviates the need for evaluating the $k_x$-plane integral for a $Q_2$ frequency — once the integral for the corresponding $Q_1$ frequency has been evaluated. Nevertheless, if one felt inclined to evaluate the $k_x$-plane integral for a $Q_2$ frequency, then we must emphasize that, in going from a $Q_1$ point $\omega$ to the corresponding $Q_2$ point $-\omega^*$, it is imperative to also switch $k_x$ and $k_z$ to $-k_x^*$ and $-k_z^*$, respectively. Thus, one must remember to flip the integration contour $C$ around the imaginary $k_x''$-axis. In the special case when the chosen $\omega$ happens to be on the positive $\omega''$-axis, the integration path taken in the $k_x$-plane could be either the contour $C$ or its flipped version around the $k_x''$-axis. As a matter of fact, on this dividing line between $Q_1$ and $Q_2$ of the $\omega$-plane, it is also allowed to directly integrate along the real $k_x'$-axis — with the accompanying benefit that it is now easy to demonstrate that the integral along the $k_x'$-axis is real-valued.

**8. Incident wavepacket**. In the preceding sections, for purposes of explaining the methodology, it sufficed to state that the incident wavepacket has a definite starting point in time and a finite footprint along the $x$-axis at the interface between media 1 and 2. For the numerical calculations that follow, we now specify a profile for the incident wavepacket arriving at the $xy$-plane at $z = 0$, as follows:

$$\boldsymbol{E}^{(\text{inc})}(x,t) = f(x/W)g(t/T)\cos(k_{xc}x - \omega_c t)\,\hat{\boldsymbol{y}}. \quad (20)$$

Here, $f(x)$ and $g(t)$, depicted schematically in Fig.6, are finite-width functions of the spatial coordinate $x$ and the time $t$, respectively. The center frequency of the packet is $\omega_c$, and its central $k_x$ is denoted by $k_{xc}$. Assuming the refractive index $n_1(\omega) = \sqrt{\varepsilon_1(\omega)}$ of the incidence medium is real-valued at $\omega = \omega_c$, the incident packet's central ray is tilted away from the $z$-axis at an angle $\theta_c = \sin^{-1}[ck_{xc}/\omega_c n_1(\omega_c)]$. Needless to say, the finite duration of the packet causes a spreading of the frequency content of the incident beam around $\omega_c$, while its finite footprint along the $x$-axis broadens the range of incidence angles around $\theta_c$, so that, in principle, every $\omega$ and every $k_x$ makes a contribution to the incident packet. A judicious choice of the envelope functions $f(x/W)$ and $g(t/T)$, however, ensures that the spatiotemporal spectrum of the incident packet remains more or less confined to the vicinity of $(k_{xc}, \omega_c)$.

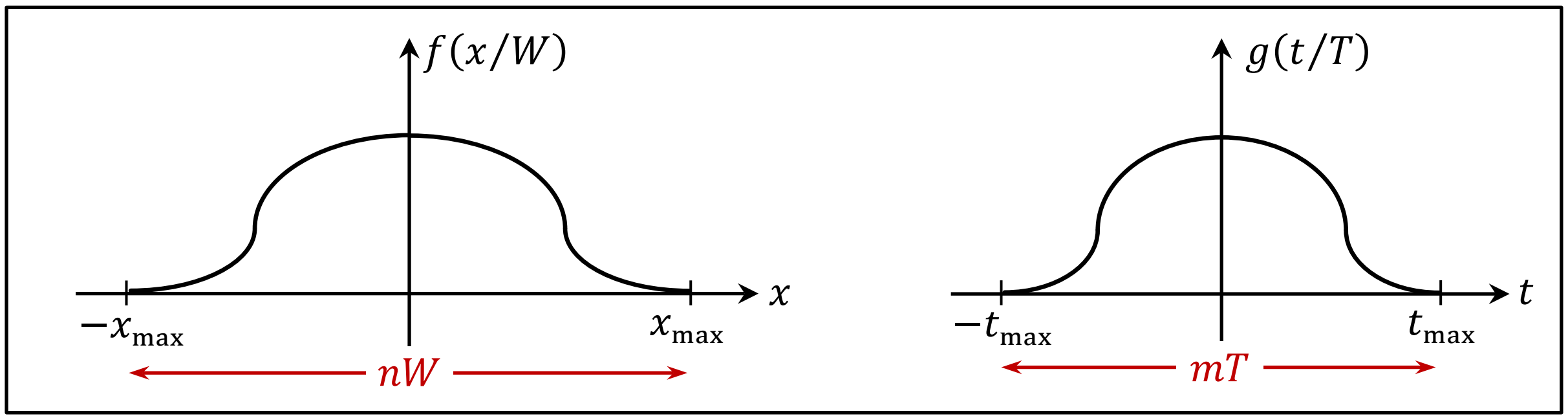

**Fig.6**. Finite-width functions $f(x/W)$ and $g(t/T)$ whose product forms the envelope of the incident wavepacket.

For the numerical simulations reported in the next section, we have chosen $f(x)$ to be a repeated convolution of the rectangular function $\text{rect}(x)$ with itself, where $\text{rect}(x)$ is defined as 1 when $|x| \le ½$ and 0 otherwise. The Fourier transform of this function, $\int_{-\infty}^{\infty} \text{rect}(x)e^{-\text{i}k_x x}\text{d}x$, is $\text{sinc}(k_x/2\pi)$, where, by definition, $\text{sinc}(s) = \sin(\pi s)/(\pi s)$. When $\text{rect}(x)$ is repeatedly convolved with itself ($n$ times), the resulting function, $\text{rect}(x) * \text{rect}(x) * \cdots * \text{rect}(x)$, will be a fairly smooth function, having width $n$ and Fourier transform $\text{sinc}^n(k_x/2\pi)$.[8] Thus, the Fourier transform of our envelope function $f(x/W)$ in Eq.(20) is given by $\tilde{f}(k_x) = W\,\text{sinc}^n(Wk_x/2\pi)$. Similarly, we have chosen the other envelope function, $g(t/T)$, so that its Fourier transform is $\tilde{g}(\omega) = T\,\text{sinc}^m(T\omega/2\pi)$. Here, $m$ and $n$ are arbitrary (albeit small) positive integers. The 2D Fourier transform of the incident packet is thus found to be

$$\tilde{E}_y^{(\text{inc})}(k_x,\omega) = ½\tilde{f}(k_x - k_{xc})\tilde{g}(\omega - \omega_c) + ½\tilde{f}(k_x + k_{xc})\tilde{g}(\omega + \omega_c). \quad (21)$$

The simple wavepacket whose spatiotemporal spectral profile is given by Eq.(21) may be criticized on the grounds that it cannot be realistically created at the interface between the incidence medium (1) and the gain medium (2), the reason being that certain frequencies are bound to be strongly absorbed within the incidence medium before arriving at the interface, and also because evanescent waves having $k_x > n_1(\omega)\omega/c$ do not survive propagation distances that are needed to reach the interface. While we concur with this critique of the model of the incident packet defined by Eqs.(20) and (21), we nevertheless contend that several important aspects of the problem of reflection from (and transmission through) gain media can be fruitfully examined by adopting such a less-than-ideal model. In Appendix A, we propose a more realistic approach to constructing the incident packet that does not suffer from the shortcomings of the model described by Eqs.(20) and (21), albeit at a significant cost in terms of the complexity of the additional numerical computations that would accompany this more realistic construction.

**9. Numerical computations**. To streamline the numerics and also to abbreviate references to numerical values of the various parameters, we use the normalization scheme summarized in Table 1. Consequently, our units of the time $t$ are $3.33\cdots$ femtoseconds (fs), frequencies $\omega$ will be specified in multiples of $3 \times 10^{14}$ radians per second (rad/s), spatial coordinates $x$ and $z$ will be in microns (µm), and the wavenumbers $k_x$ and $k_z$ will be given in units of $10^6$ inverse meters ($\mathrm{m}^{-1}$). Note that this normalization scheme leaves products such as $\omega t$, $k_x x$, and $k_z z$ intact. Normalized entities are identified by an overbar; thus, $\bar{\omega} = 10$ represents the angular frequency $\omega = 3 \times 10^{15}$ rad/s, and $\bar{t} = 6$ is the normalized time at $t = 20$ fs.

| $t = t_0\bar{t}$ | $\omega = \omega_0\bar{\omega}\quad (\omega_0 = 1/t_0)$ | $x = x_0\bar{x}$ | $k_x = k_0\bar{k}_x\quad (k_0 = 1/x_0)$ |
|---|---|---|---|
| $t_0 = 3.33 \times 10^{-15}$ s | $\omega_0 = 3.0 \times 10^{14}$ rad/s | $x_0 = 1.0 \times 10^{-6}$ m | $k_0 = 1.0 \times 10^{6}$ 1/m |

**Table 1**. The time $t$, angular frequency $\omega$, distance $x$, and wavenumber $k_x$ are normalized by $t_0$, $\omega_0$, $x_0$, and $k_{x0}$.

Each of the three media (incidence, gain, and transmittance) is modeled using a single Lorentz oscillator (that is, referencing Eq.(9), $K = 1$). For each oscillator, the plasma frequency $\omega_p$, the resonance frequency $\omega_r$, and the damping coefficient $\gamma$ have the numerical values listed in Table 2. The incidence and transmittance media, being partial absorbers in the vicinity of their respective resonance frequencies, have oscillator strengths $f_1 = f_3 = +1$, whereas medium 2, the gain medium, has $f_2 = -1$. For the finite-slab computations pertaining to Fig.1(b), we have chosen the thickness $d$ of the gain medium to be 5 µm.

| | | | | |
|---|---|---|---|---|
| Incidence (1) | $\omega_{p1} = 1.5 \times 10^{15}$ | $\omega_{r1} = 4.5 \times 10^{15}$ | $\gamma_1 = 3.0 \times 10^{14}$ | |
| | $\bar{\omega}_{p1} = 5.0$ | $\bar{\omega}_{r1} = 15.0$ | $\bar{\gamma}_1 = 1.0$ | |
| Gain (2) | $\omega_{p2} = 3.3 \times 10^{14}$ | $\omega_{r2} = 3.0 \times 10^{15}$ | $\gamma_2 = 3.0 \times 10^{14}$ | $d = 5$ µm |
| | $\bar{\omega}_{p2} = 1.1$ | $\bar{\omega}_{r2} = 10.0$ | $\bar{\gamma}_2 = 1.0$ | $\bar{d} = 5.0$ |
| Transmittance (3) | $\omega_{p3} = 1.2 \times 10^{15}$ | $\omega_{r3} = 4.8 \times 10^{15}$ | $\gamma_3 = 3.0 \times 10^{14}$ | |
| | $\bar{\omega}_{p3} = 4.0$ | $\bar{\omega}_{r3} = 16.0$ | $\bar{\gamma}_3 = 1.0$ | |

**Table 2**. Material parameters for the incidence medium (1), gain medium (2), and transmittance medium (3). When the gain medium is not semi-infinite, its thickness is specified as $d$. Also shown are normalized parameter values.

The incident wavepacket is modelled using Eq.(20), with the spatiotemporal envelopes depicted in Fig.6 and the numerical parameter values listed in Table 3. Thus, the entire duration of the incident pulse is $mT = 40$ fs (i.e., in normalized units, the pulse arrives at $\bar{t} = -6$ and terminates at $\bar{t} = 6$), while the total footprint of the beam is $nW = 25$ µm along the $x$-axis. The central vacuum wavelength of the incident packet is $\lambda_c = 2\pi c/\omega_c = 0.754$ µm, while the central ray's incidence angle is $\theta_c \cong 63°$. Considering that $n_1(\omega_c) \cong 1.075$ and $n_2(\omega_c) \cong 1.0$, this $\theta_c$ is only slightly less than the critical angle of total internal reflection at the interface between media 1 and 2. (At $\omega = \omega_c$, the critical angle is $\theta_{\mathrm{TIR}} = \sin^{-1}(n_2/n_1) \cong 68.5°$.)

| | | | | | |
|---|---|---|---|---|---|
| $\omega_c = 2.5 \times 10^{15}$ | $T = 1.0 \times 10^{-14}$ | $m = 4$ | $k_{xc} = 8.0 \times 10^{6}$ | $W = 5.0 \times 10^{-6}$ | $n = 5$ |
| $\bar{\omega}_c = 8.3333$ | $\bar{T} = 3.0$ | | $\bar{k}_{xc} = 8.0$ | $\bar{W} = 5.0$ | |

**Table 3**. The incident wavepacket is specified by Eq.(20) and its spatiotemporal envelopes are depicted in Fig.6. The pulse duration is $mT$, the footprint of the beam is $nW$, its center frequency is $\omega_c$, and its central $k_x$ value is $k_{xc}$. Also shown are the normalized values of these parameters.

We have chosen these particular material parameters and incident wave profile for two reasons: (i) to allow investigation of the TIR regime, while (ii) keeping the computation time manageable. (We discuss later how the computation time is affected by the material parameters.)

Plots of the real and imaginary parts of the complex refractive index $n(\omega) = \sqrt{\varepsilon(\omega)}$ for media 1, 2, and 3 are shown in Fig.7. Considering that the central frequency of the incident packet is $\bar{\omega}_c = 8.3333$, both the incidence and transmittance media are seen to be highly transparent for a good fraction of the frequency content of the incident beam, which is below $\bar{\omega} \cong 11$. The gain coefficient of medium 2 peaks at $\bar{\omega} = 10$, thus allowing for significant overlap between the incident frequencies and those that are amplified by the gain medium.

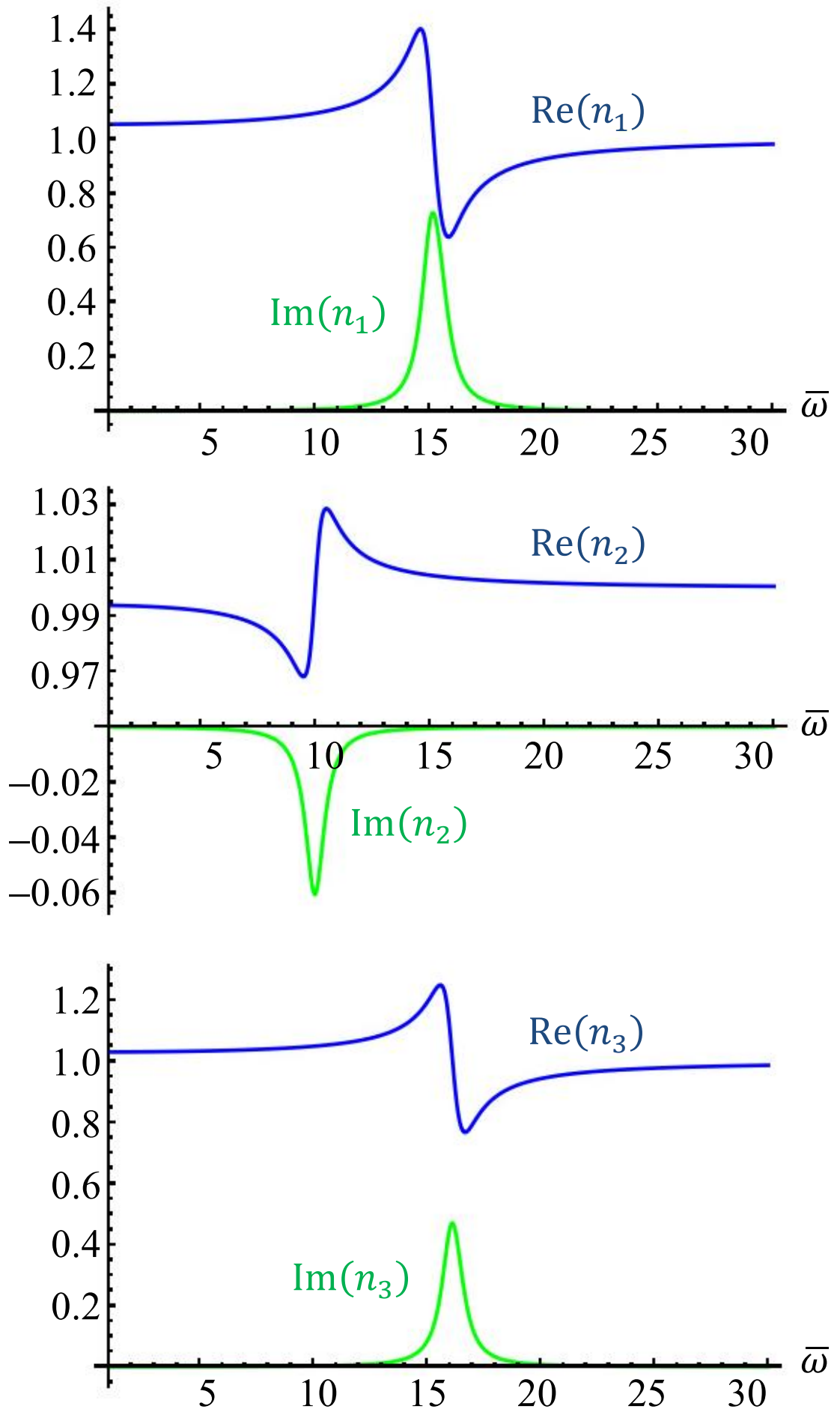

**Fig. 7**. Plots of the complex refractive indices $n_1(\omega)$, $n_2(\omega)$, $n_3(\omega)$ versus the frequency $\omega$ of the exciting radiation. By definition, $n(\omega) = \sqrt{1 \pm \omega_p^2/(\omega_r^2 - \omega^2 - \mathrm{i}\gamma\omega)}$, where the plus sign is used for the incidence and transmittance media (1 and 3), and the minus sign for the gain medium (2).

Our next task is to identify the contour $C$ of integration within the $k_x$-plane. In the case of the finite-thickness slab of Fig.1(b), we start by selecting a very large, randomly-distributed set of points inside a rectangular region below the real axis in the $k_x$-plane; the number of points in this rectangular region is typically in the millions. Each of these points is then used as an initial guess for a Newton-based numerical pole finder, where the poles are solutions of $\nu(k_x, \omega) = 1$; see Eq.(16). Many guesses lead to no poles, and a single pole can be determined by many different guesses. In this way, we find a distribution of poles in $Q_4$ of the $k_x$-plane such as that displayed in Fig.8(a). We have verified that the general pattern of the pole distribution and, most importantly, the spread of the poles below the real axis and down into $Q_4$, do not change when we repeat the calculation using different number of points and different rectangular regions.

The blue dots in Fig.8(a) are the poles of the Fresnel reflection (or transmission) coefficient corresponding to real and positive values of $\omega$ that land in $Q_4$ of the $k_x$-plane. Starting at a fixed $\omega'$ in the $\omega$-plane and moving up parallel to the imaginary $\omega''$-axis, the corresponding blue dots in Fig.8(a) are found to move upward, as depicted in Figs.8(b, c), where, for clarity, we have plotted only a subset of all the pole trajectories. The integration contour (green) is subsequently chosen to bypass all possible poles corresponding to (complex) frequencies in $Q_1$ of the $\omega$-plane. (A flipped copy of the same contour in $Q_2$ of the $k_x$-plane is eventually needed to complete the integration path.)

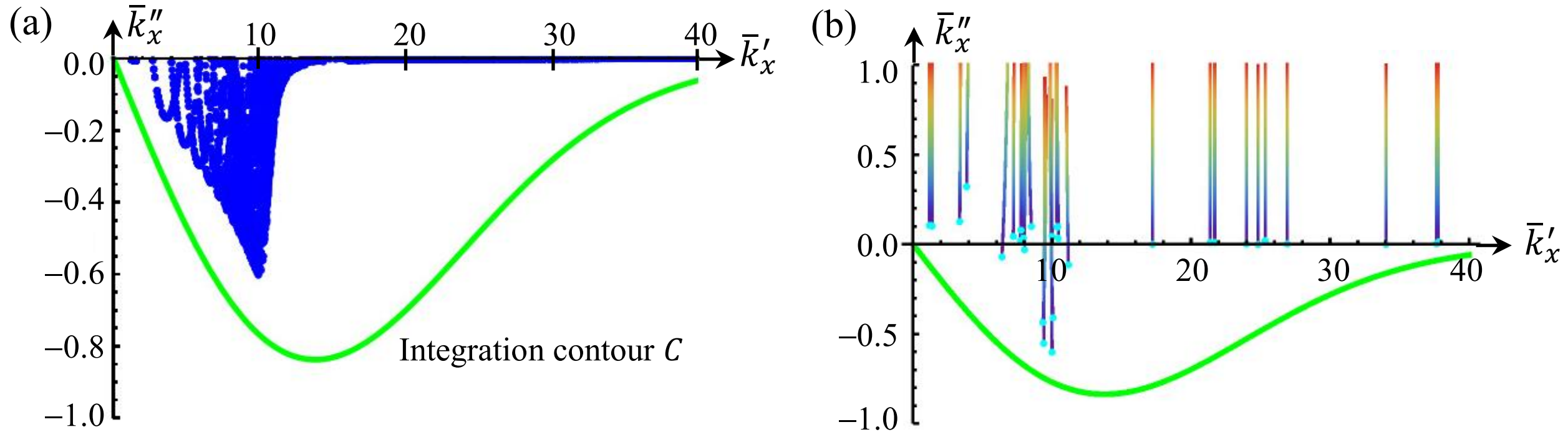

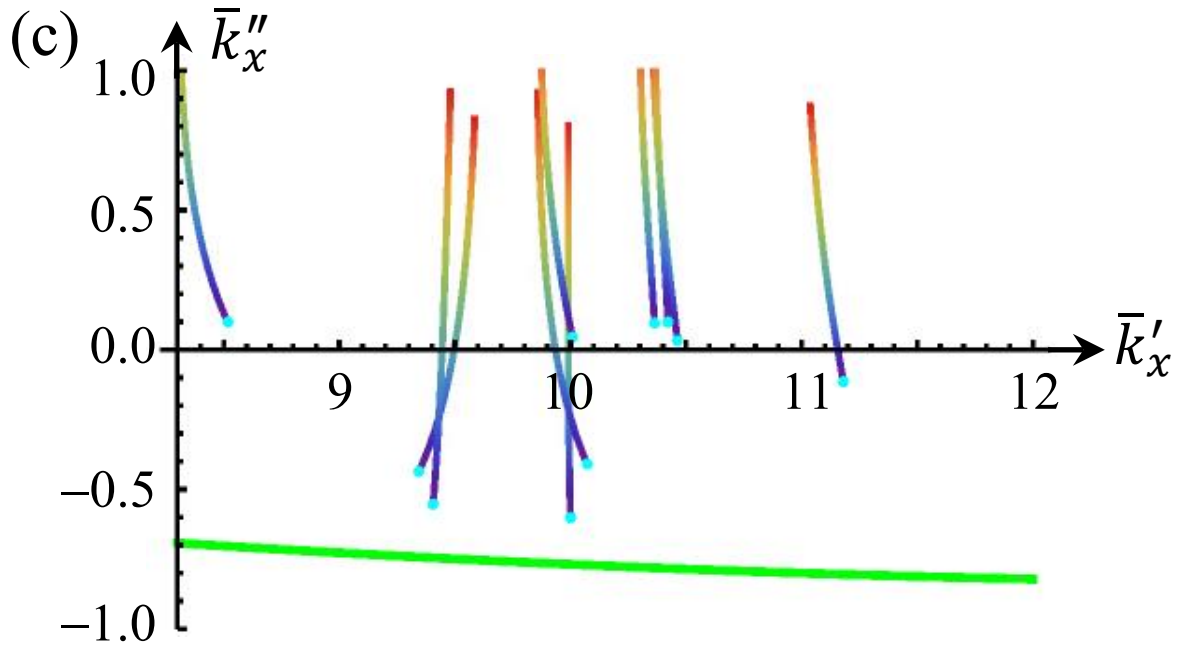

**Fig.8**. Integration contour (green) in $Q_4$ of the $k_x$-plane. A flipped copy of the same contour in $Q_2$ is also needed to complete the integration path. (a) The blue dots are the poles of the Fresnel reflection (or transmission) coefficient for a large number of points $\omega'$ on the positive real frequency axis. (b) Trajectories of a few randomly-selected poles in the $k_x$-plane when the corresponding frequency $\omega$ moves up from $\omega'$ to $\omega' + \mathrm{i}\omega''$; the purple-to-red color coding indicates increasing values of $\bar{\omega}''$ from 0 to 1. (c) Magnified view of a small section of the $k_x$-plane, showing the region $8.3 \leq \bar{k}_x' \leq 12$ and $-1 \leq \bar{k}_x'' \leq 1$.

Lastly, we need to verify that no poles further down in $Q_4$ of the $k_x$-plane have escaped our attention. It can be shown with an asymptotic approximation to the pole equation $\nu(k_x, \omega) = 1$, valid far from the real axis $k_x'$, that the poles corresponding to complex frequencies in $Q_1$ of the $\omega$-plane are bounded from below in $Q_4$ of the $k_x$-plane—i.e., there exists a value of $k_x''$ below which no poles can be found. Therefore, we have confidence that our numerical method has found all of the pole trajectories, and that our chosen contour $C$ avoids crossing any of them.

We have previously mentioned that the branch-cuts of $k_{1z}$ and $k_{3z}$ do not cross the $k_x'$-axis and, therefore, would not by themselves require deformation of the standard contour along the $k_x'$-axis. Figure 9(a) illustrates this point for $k_{1z}$, showing that, for $\omega' \geq 0$, the branch-points for the passive semi-transparent medium 1 always land in $Q_1$ (and, by symmetry, in $Q_3$) of the $k_x$-plane. The situation is different, however, for the gain medium 2, since, for $\omega' \geq 0$, some of the $k_{2z}$ branch-points originate in $Q_4$ (and, by symmetry, also in $Q_2$) of the $k_x$-plane; see Fig.9(b). Now, for the finite-slab problem of Fig.1(b), the branch-cuts of $k_{2z}$ are inconsequential, since Eqs.(13)-(18) are indifferent to a sign-change of $k_{2z}$. It is only for the single-surface problem of Fig.1(a) that the $k_{2z}$ branch-cuts need to be considered when constructing the integration contour $C$ in the $k_x$-plane. Figure 9(b) shows an integration contour (green) that is properly chosen to stay below all $k_{2z}$ branch-cuts in $Q_4$ of the $k_x$-plane. (Not shown in this figure is the other half of the same contour $C$ that is symmetrically located in $Q_2$ of the $k_x$-plane.) The general rule is that the $k_x$-plane integration contour $C$ must avoid crossing all branch-point trajectories as well as all the pole trajectories (if any) corresponding to frequencies $\omega$ in $Q_1$ of the $\omega$-plane; this is what Figs.8 and 9 aim to convey. Exceptions arise in the case of finite-slabs, where $k_{2z}$ branch-points become inconsequential, and in the single-surface problem, where $k_{3z}$ is non-existent.

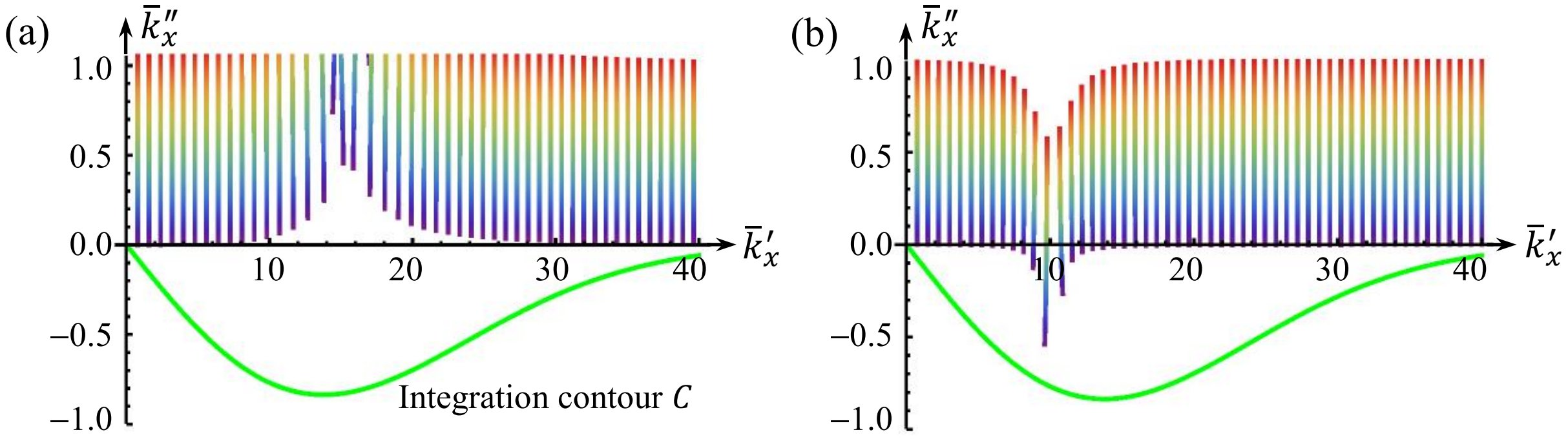

**Fig.9**. Branch-point trajectories of (a) $k_{1z}$ and (b) $k_{2z}$, corresponding to a large number of complex frequencies $\omega = \omega' + \mathrm{i}\omega''$, with $0 \leq \bar{\omega}' \leq 40$ and $\bar{\omega}''$ rising from 0 to 1 (purple to red). Also shown (in green) is the integration contour in $Q_4$ of the $k_x$-plane, which is used in our numerical evaluation of the inverse Fourier integrals over $k_x$. For the parameter values used here, the branch-point trajectories appear to be vertical lines at the scale of the plot, but zooming in would reveal their slight curviness.

As a reminder of the role played by the branch-points, branch-cuts, and the contour $C$ in our numerical simulations, Fig.10(a) shows that the location of a point $k_x$ on $C$ determines the two complex numbers on the right-hand side of Eq.(19) that are needed to identify unique values for $k_{1z}$, $k_{2z}$, and $k_{3z}$. The indicated ranges for the angle $\varphi$ associated with each branch-cut guarantee that all three $k_z$s approach $\omega/c$ in the limit when $\omega'' \to +\infty$, which is essential for the satisfaction of the causality constraint. Note that the chosen branch-cuts do *not* cross the contour $C$, which is also necessary to ensure the analyticity of the integrands in Eqs.(7) and (8) by making the angles $\varphi$ vary smoothly as the point $k_x$ travels along $C$.

As an example, consider Fig.10(b), which shows the trajectory of $k_{2z} = k_{2z}' + \mathrm{i}k_{2z}''$ at the fixed frequency $\bar{\omega} = 10$, when $k_x$ moves along $C$ from $\bar{k}_x' = -30$ to 30. It is seen that, for $|\bar{k}_x'| < 8.76$, the imaginary part $k_{2z}''$ of $k_{2z}$ is negative, indicating that the corresponding plane-wave inside the semi-infinite gain medium 2 exponentially grows along the $z$-axis. Outside this interval, where $|\bar{k}_x'| > 8.76$, the plane-wave inside the gain medium decays away from the interfacial plane at $z = 0$. It is thus seen that the problem of TIR at the interface between a (nearly) transparent incidence medium 1 and a (weakly amplifying) semi-infinite gain medium 2 is not amenable to an elementary solution; rather, it is essential to consider the entire spatio-temporal spectrum of the incident packet to determine which plane-wave constituents of the incident beam participate in forward amplification upon entering the gain medium, and which ones support the backward propagating (and amplified) reflected beam.

The forward and inverse Fourier transforms needed in our numerical computations are given by

$$\tilde{E}_y(k_x, \omega) = \int_{-\infty}^{\infty} \mathrm{d}t \int_{-\infty}^{\infty} E_y(x,t) e^{-\mathrm{i}(k_x x - \omega t)} \mathrm{d}x. \tag{22}$$

$$E_y(x,t) = (2\pi)^{-2} \int_{-\infty}^{\infty} \mathrm{d}\omega \int_C \tilde{E}_y(k_x, \omega) e^{\mathrm{i}(k_x x - \omega t)}\, \mathrm{d}k_x. \tag{23}$$

If we now scale the space and time coordinates, the frequency, the wave-number, the field amplitude, and the transformed field amplitude, we will have

$$\tilde{E}_y(\bar{k}_x, \bar{\omega}) = E_0 x_0 t_0 \int_{-\infty}^{\infty} \mathrm{d}\bar{t} \int_{-\infty}^{\infty} \bar{E}_y(\bar{x}, \bar{t}) e^{-\mathrm{i}(\bar{k}_x \bar{x} - \bar{\omega}\bar{t})} \mathrm{d}\bar{x}. \tag{24}$$

Suppose the incident $E$-field amplitude is scaled such that $E_0 x_0 t_0 = 1$. A direct inverse Fourier transformation of Eq.(24) then yields

$$\bar{E}_y(\bar{x}, \bar{t}) = (2\pi)^{-2} \int_{-\infty}^{\infty} \mathrm{d}\bar{\omega} \int_C \tilde{E}_y(\bar{k}_x, \bar{\omega}) e^{\mathrm{i}(\bar{k}_x \bar{x} - \bar{\omega}\bar{t})} \mathrm{d}\bar{k}_x. \tag{25}$$

It should be clear that the inverse Fourier transformation can be carried out with the normalized (or scaled) values of $x$, $t$, $k_x$, and $\omega$. Moreover, the proportionality of the reflected (or transmitted) wavepacket amplitude to that of the incident packet guarantees that the choice of the scale factor $E_0$ for the $E$-field amplitude is inconsequential. Thus, Eq.(25) is the fundamental equation in our inverse Fourier transform calculations.

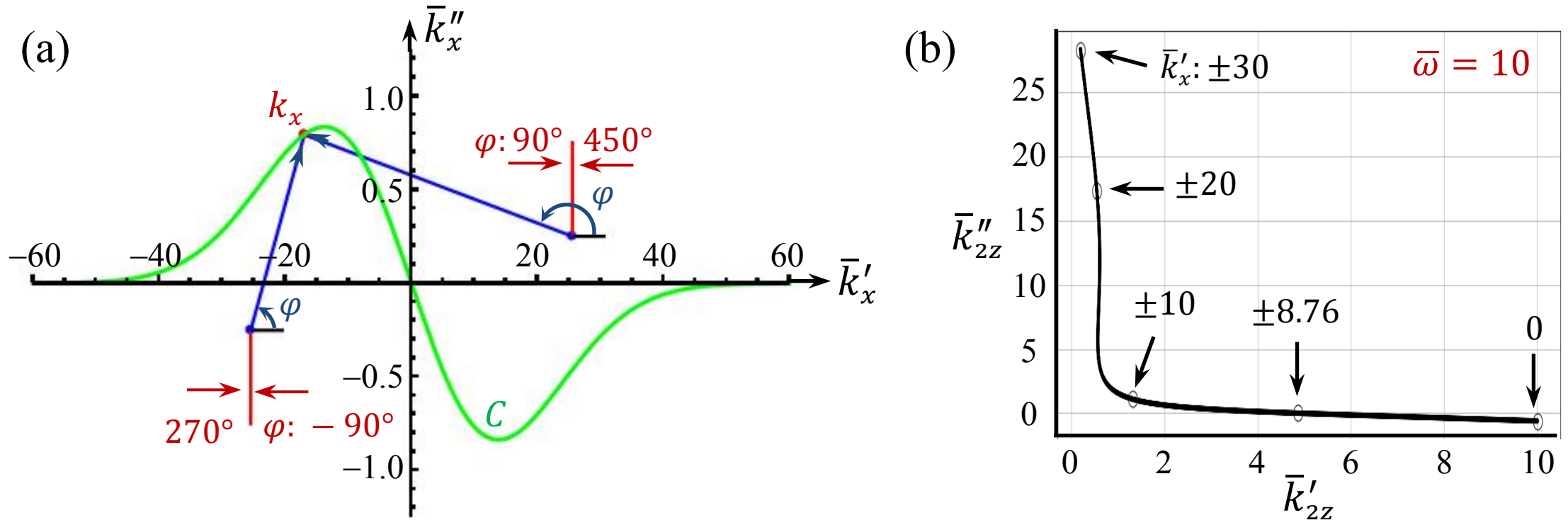

**Fig. 10**. (a) Integration contour (green) in the $k_x$-plane, together with a pair of branch-cuts (red) corresponding to $k_{1z}$ at $\bar{\omega} = 26 + 0.2\mathrm{i}$. For this (arbitrarily chosen) point in $Q_1$ of the $\omega$-plane, the branch-points $k_x = \pm\omega n_1(\omega)/c$ are at $\bar{k}_x = \pm(25.5 + 0.25\mathrm{i})$. Numerically, the $k_x$-integral is evaluated over the entire contour $C$ for a fixed value of $\omega$, then repeated for each $\omega$ on the positive $\omega'$-axis. At each point $k_x$ along $C$, we draw the two blue arrows from the branch-points to $k_x$; these arrows represent the two terms whose complex square roots appear on the right-hand side of Eq.(19). The phase $\varphi$ of each of these complex numbers falls within the 360° range specified at the corresponding branch-cut. In this way, the complex square roots are unambiguously computed, yielding a unique value for $k_{1z}$. A similar procedure is used to uniquely specify the values of $k_{2z}$ and $k_{3z}$. (b) Trajectory of $\bar{k}_{2z} = \bar{k}_{2z}' + \mathrm{i}\bar{k}_{2z}''$ corresponding to $\bar{\omega} = 10$, as $k_x$ moves along $C$ from $\bar{k}_x' = -30$ to 0. Subsequently, $\bar{k}_{2z}$ returns along the same trajectory as $\bar{k}_x'$ continues to rise from 0 to 30.

Figures 11-14 pertain to the single-surface problem involving a semi-infinite gain medium. The geometric configuration of the simulated system is depicted in Fig.11. The incident packet is a 40 fs linearly-polarized (TE) light pulse with a 25 µm footprint, arriving at the interface between the passive medium 1 and the semi-infinite gain medium 2 at an oblique angle of ~63°. The incident light pulse is confined to the (normalized) time interval $[\bar{t}_{\min}, \bar{t}_{\max}] = [-6, 6]$.

In Fig.12, we show the reflected wavepackets at several locations within the interfacial plane (i.e., at $z = 0$); also shown for comparison are the incident packets at $x_0 = 0$ and ±5 µm. The reflected $E$-field amplitude profiles are seen to be broadened (due to dispersion as well as diffraction), and also proportionately delayed with an increasing distance from the incident beam's center at $(x, y, z) = (0, 0, 0)$. Although the reflected waves in the vicinity of the incident beam are relatively weak, they gain strength with an increasing distance $|x_0|$ away from the center. The tilt of the incident beam is such that its spatial frequency content is heavily biased in favor of positive $k_x$ values. Nevertheless, plane-waves having negative $k_x$ values are also present within the spectral profile of the incident wavepacket; these backward-propagating waves (along the $x$-axis) are responsible for the growth of the reflected wave amplitudes along the negative $x$-axis. Causality is seen to be satisfied since the reflected pulses everywhere emerge only after a proper delay following the onset of the incident pulse at $\bar{t} = -6$.

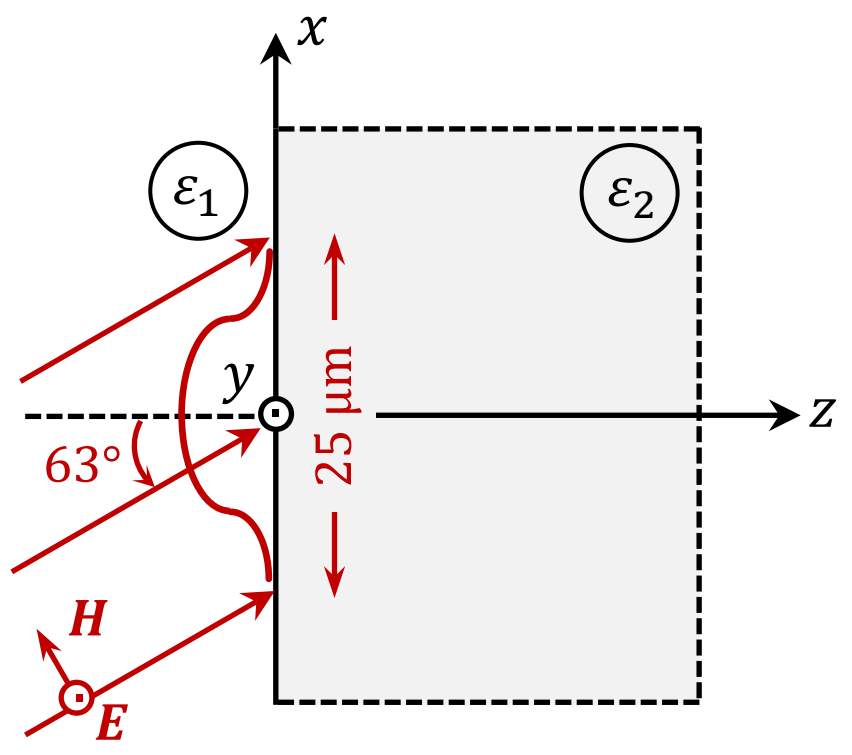

**Fig.11**. A 40 fs wavepacket, linearly polarized along the $y$-axis and having a 25 μm footprint along the $x$-axis, arrives at the interface between a passive, nearly-transparent medium 1 and a weakly-amplifying, semi-infinite medium 2. The central ray of the incident packet makes an angle $\theta_c \cong 63°$ with the $z$-axis. Plots of the refractive indices $n_1 = \sqrt{\varepsilon_1}$ and $n_2 = \sqrt{\varepsilon_2}$ as functions of the temporal frequency $\omega$ appear in Fig.7.

Figure 13 shows the reflected packets at a few locations within the $xy$-planes at $z = -5$ μm and $z = -10$ μm, which are several wavelengths away from the interfacial plane at $z = 0$. Recalling that the central incident ray arrives at the rather large oblique angle of $\theta_c \cong 63°$, we note that the reflected central ray is merely 27° or so away from the interfacial plane. Thus, for example, the reflected packets at $(x_0, z_0) = (20, -5\ \mu\text{m})$ and $(20, -10\ \mu\text{m})$ could just be the delayed (and possibly attenuated) versions of the reflected pulses that leave the interface at $(x_0, z_0) = (10\ \mu\text{m}, 0)$ and $(0, 0)$, respectively; see Fig.12. Considering that the packet arriving at $(20, -5\ \mu\text{m})$ has nearly twice the amplitude of that at $(20, -10\ \mu\text{m})$, it is likely that the reduced amplitude at the latter location, being only one-fifth of that at $(0, 0)$, is caused by diffraction or by interference with other parts of the reflected light that return from the interfacial plane.

The transmitted waves at several points inside the gain medium are shown in Fig.14. Note that the transmitted packets, in addition to being broadened and properly delayed, are also substantially amplified. The arrival time of the transmitted packet at $(x_0, z_0) = (0, 30\ \mu\text{m})$ is consistent with the expected minimum delay of $\Delta\bar{t} = 30$. Similarly, within the $xy$-plane at $z = 40$ μm, the pulses arrive a short while after the expected minimum delay of $\Delta\bar{t} = 40$.

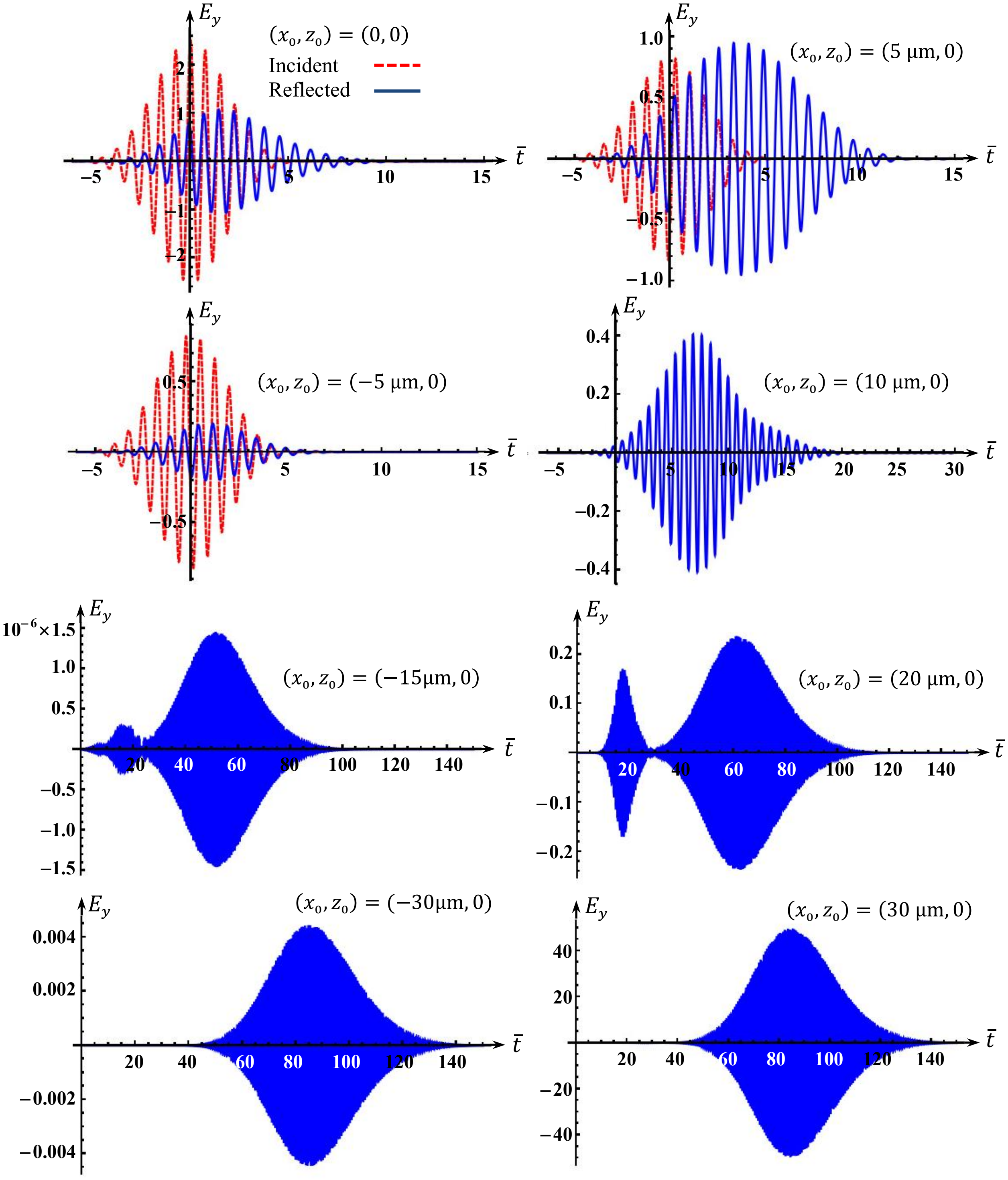

**Fig.12**. Plots of the incident (red) and reflected (blue) wavepackets at various locations in the interfacial plane (i.e., at $z = 0$) between the nearly-transparent medium 1 and the weakly-amplifying, semi-infinite medium 2. On the right-hand side, the reflected $E$-field packet is seen to rapidly grow with the positive distance $x_0$ away from the incident beam's footprint. As for negative values of $x_0$ appearing on the left-hand side, the $E$-field amplitude is small at first, but it also begins to rise, albeit slowly, with distance from the incident packet's center at $(x, y, z) = (0,0,0)$. The packets are seen to be broadened and also proportionately delayed while gaining strength as the observation point recedes from the incident beam's footprint.

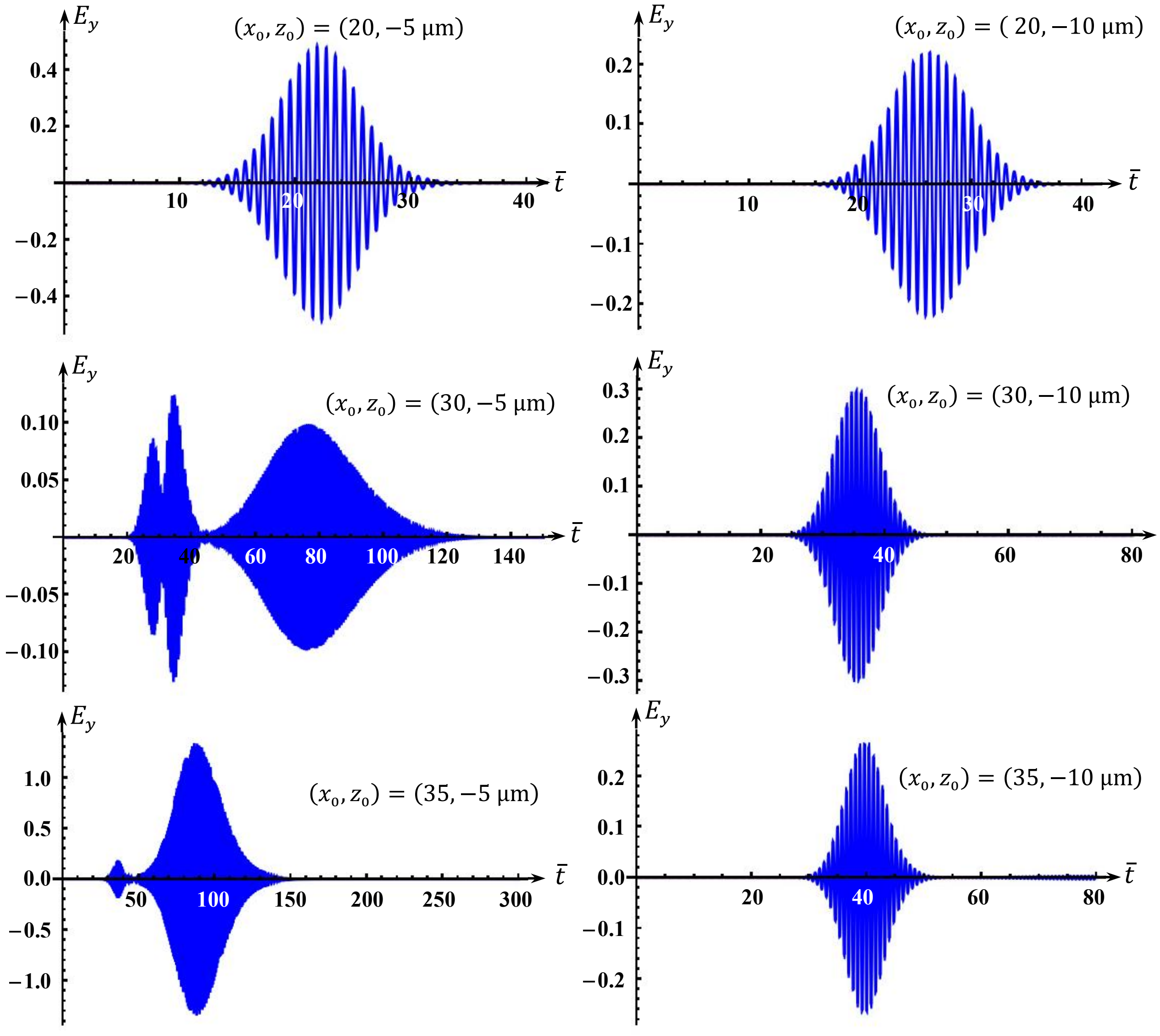

**Fig.13**. Reflected wavepackets within the $xy$-planes located at $z = -5\ \mu\text{m}$ (left column) and $z = -10\ \mu\text{m}$ (right column). From top to bottom, the distance along the $x$-axis from the center of the incident beam at $(x, y, z) = (0, 0, 0)$ is $x_0 = 20,\ 30,\ 35\ \mu\text{m}$. The packets are seen to be broadened (due to dispersion as well as diffraction) and also proportionately delayed with distance away from the incident beam's footprint.

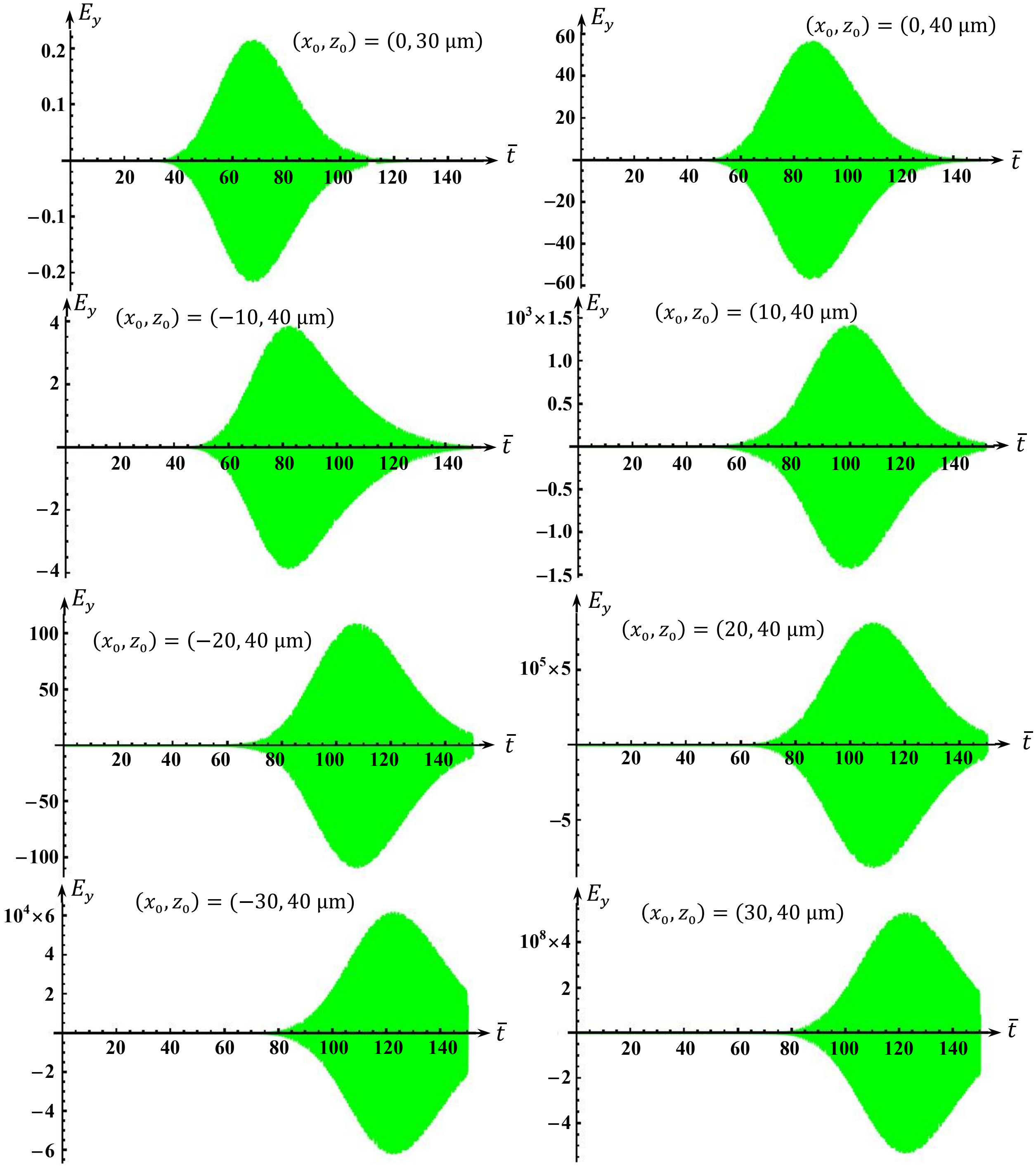

**Fig.14**. Transmitted packets inside the gain medium at various distances from the center $(x, y, z) = (0,0,0)$ of the incident beam. From top to bottom, the column on the left-hand side corresponds to $x_0 = 0, -10, -20, -30$ μm, whereas the column on the right represents the points $x_0 = 0, 10, 20, 30$ μm. The packets are seen to be broadened and proportionately delayed with distance away from the incident beam's footprint; they are also significantly amplified, especially on the right-hand side, where $x_0$ is positive.

Our next set of numerical results pertains to reflected and transmitted packets for a 5 μm-thick gain medium (2) sandwiched between the incidence medium (1) and the transmittance medium (3). The geometrical configuration of the system is shown in Fig.15. The incident packet is, once again, a 40 fs linearly-polarized (TE) light pulse with a 25 μm footprint, which arrives at the interface between the passive medium 1 and the 5 μm-thick gainy slab 2 at an oblique angle of $\sim 63°$. The light passing through the slab emerges into the nearly-transparent passive medium 3.

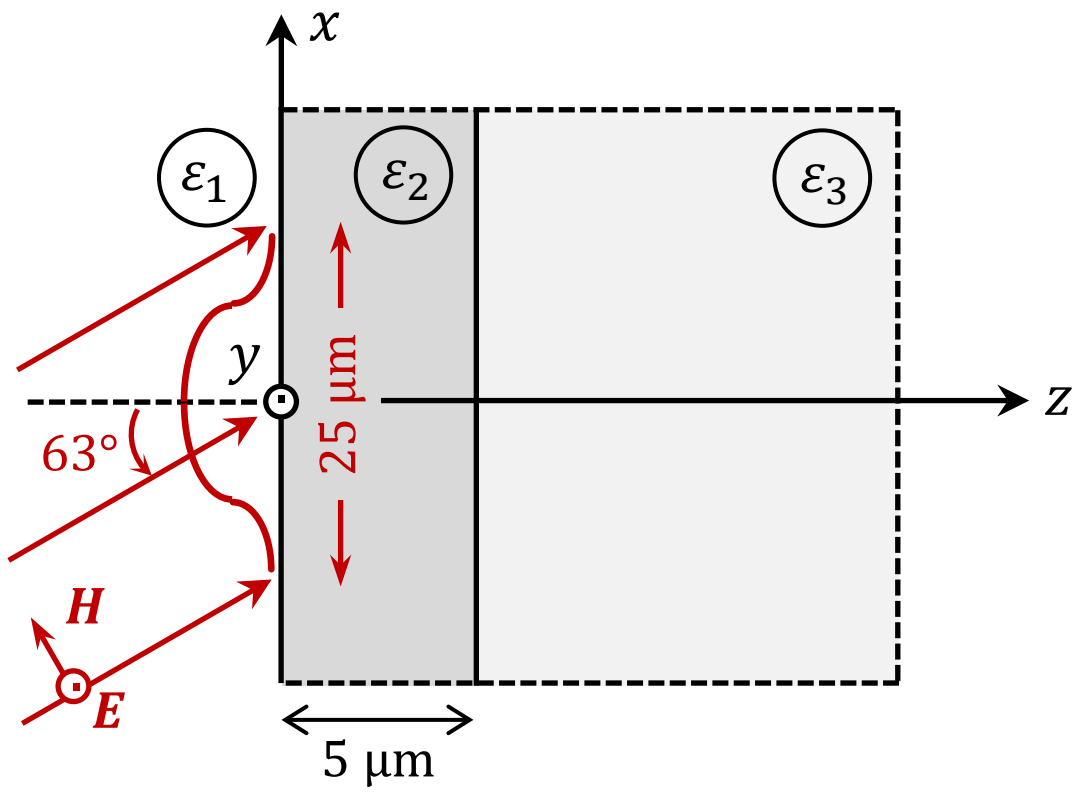

**Fig.15**. A 40 fs wavepacket, linearly-polarized along the $y$-axis and having a 25 μm footprint along the $x$-axis, arrives at the interface between a passive, nearly-transparent medium 1 and a weakly-amplifying medium 2. The light passing through the 5 μm-thick gainy slab emerges into another nearly-transparent, semi-infinite, passive medium 3. The central ray of the incident packet makes an angle $\theta_c \cong 63°$ with the $z$-axis. Plots of the complex refractive indices $n_1 = \sqrt{\varepsilon_1}$, $n_2 = \sqrt{\varepsilon_2}$, and $n_3 = \sqrt{\varepsilon_3}$ as functions of the temporal frequency $\omega$ appear in Fig.7.

Figure 16 shows the reflected wavepackets at several locations within the $xy$-plane at $z = 0$ (i.e., at the interface between the nearly-transparent medium 1 and the 5 μm-thick amplifying medium 2); also shown, for comparison with the reflected waves, are the incident packets at $x_0 = 0, \pm 5$ μm. The incident pulse, centered at the entrance facet of the slab at $(x, y, z) = (0,0,0)$, is confined to the (normalized) time interval $[\bar{t}_{\min}, \bar{t}_{\max}] = [-6, 6]$. The reflected $E$-field amplitude profiles are seen to be broadened, due to dispersion as well as diffraction, and also properly delayed relative to the onset of the incident pulse at $\bar{t} = -6$, the delay confirming the causality of the system's response. In some cases, the pulse that is directly reflected from the front facet of the slab can be distinguished from that returning from the rear facet. In other cases, a first pulse, having a different spectral content than a second pulse, travels at a greater group velocity inside the slab and arrives at the observation point prior to the emergence of the slower second pulse. As expected, in cases where the first reflected pulse originates at the front facet of the slab (e.g., at $x_0 = 0, \pm 5$ μm, 10 μm), a comparison with Fig.12 shows that it coincides with the reflected pulse from the semi-infinite gain medium.

The negative spatial frequency content of the incident packet (corresponding to incident plane-waves having negative $k_x$ values) generates a backward-propagating wave along the negative $x$-axis, which gains strength as it moves away from the incident packet's footprint. The exponential growth of this backward wave that travels along the negative $x$-axis, as well as that of the forward-propagating wave along the positive $x$-axis, are manifestations of the so-called "convective instability" associated with such weakly-amplifying slabs.[11-14]

Figure 17 shows several transmitted packets emerging from the rear facet of the gainy slab at $z = 5$ μm. In general, these packets are broadened, properly delayed with distance away from the incident beam's footprint, and amplified in consequence of single or multiple passages through the slab. Occasionally, one can distinguish the light that directly reaches the rear facet from that which requires an extra bounce inside the slab.

The computed results reported here were obtained using 1000 points along the integration contour in the $k_x$-plane $(-40 \le \bar{k}'_x \le 40)$; the number of points on the positive real $\omega'$-axis $(0 \le \bar{\omega}' \le 40)$ was 2000 for the semi-infinite gain medium and 4000 for the 5 μm-thick slab. Numerical accuracy was achieved using 100 working digits, with 15 digits as the precision goal.

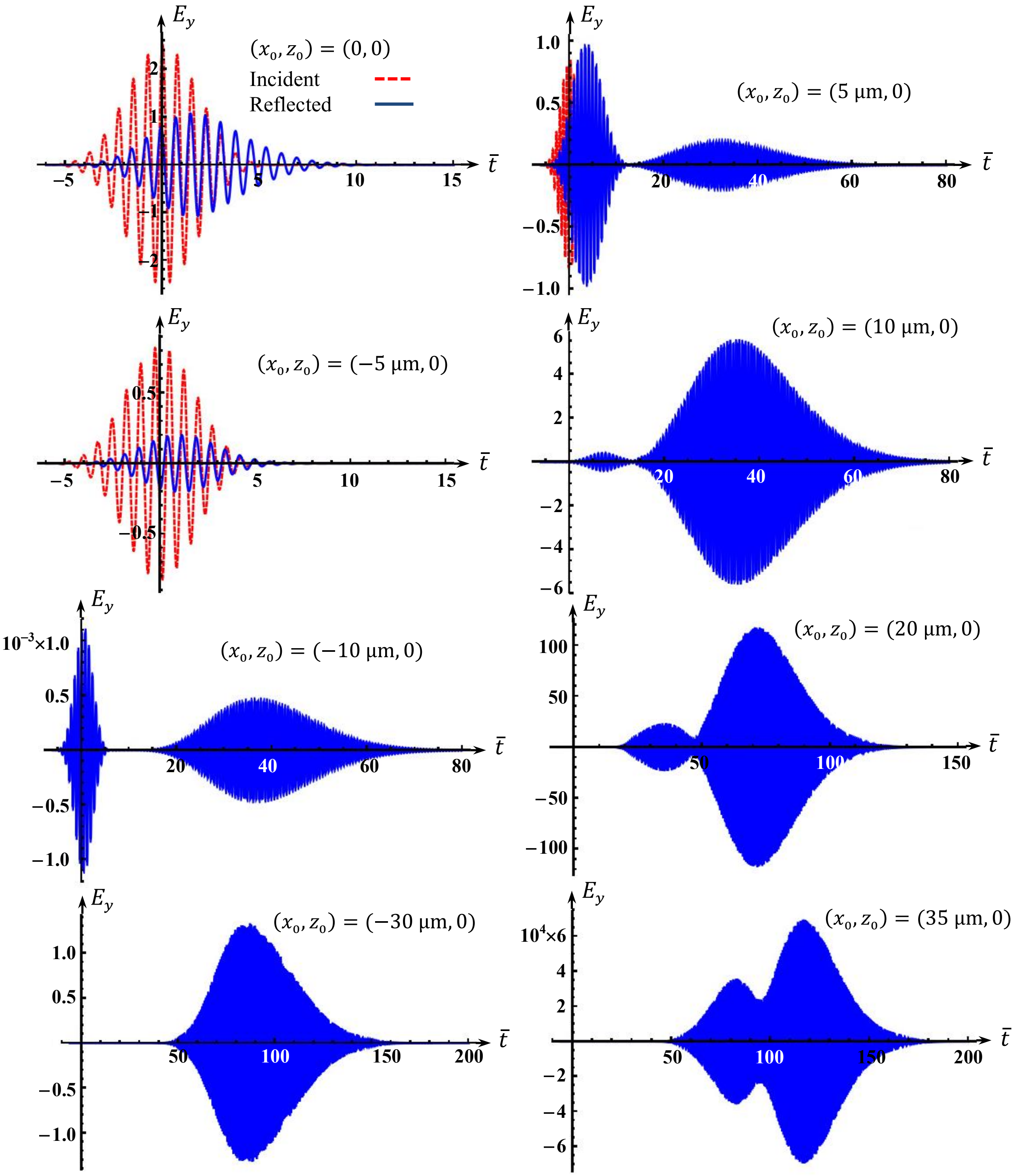

**Fig.16**. Plots of the incident (red) and reflected (blue) wavepackets at several locations in the interfacial plane (i.e., the $xy$-plane at $z = 0$) between the nearly transparent medium 1 and the 5 μm-thick weakly amplifying medium 2. On the right-hand side, the reflected $E$-field amplitude is seen to rapidly increase with the positive distance $x_0$ away from the center of the incident packet. As for the negative values of $x_0$ appearing on the left-hand side, the $E$-field amplitude is fairly small at first, but it also rises, albeit slowly, with the increasing distance from the incident packet's center at $(x, y, z) = (0,0,0)$. The packets are broadened and proportionately delayed as the observation point recedes from the incident wave's footprint.

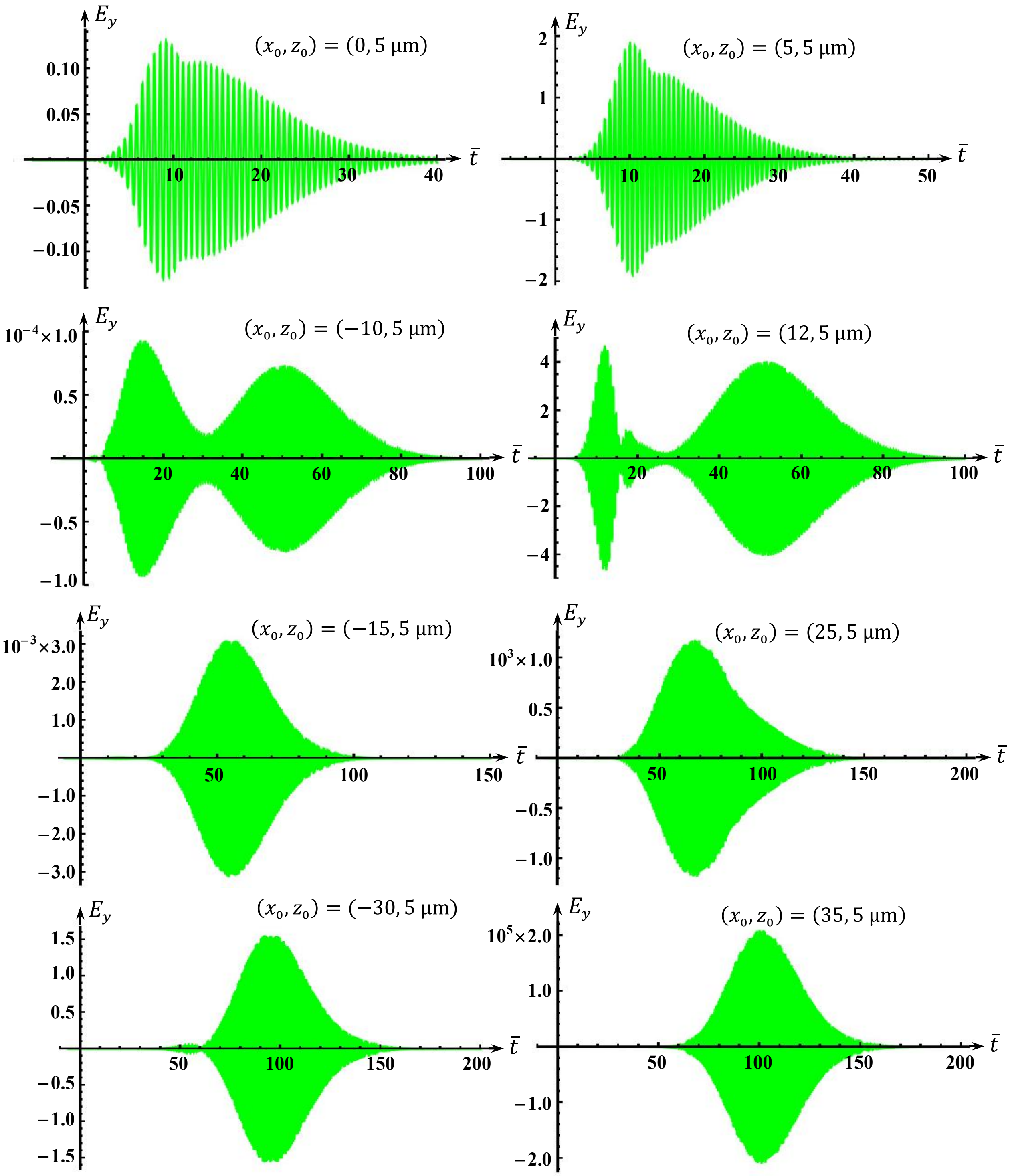

**Fig.17**. Transmitted packets emerging from several locations at the rear facet of the gainy slab at $z = 5\ \mu\text{m}$. From top to bottom, the column on the left-hand side corresponds to $x_0 = 0, -10, -15, -30\ \mu\text{m}$, whereas the column on the right represents the points $x_0 = 5,\ 12,\ 25,\ 35\ \mu\text{m}$. The packets are seen to be broadened and proportionately delayed with distance away from the incident beam's footprint. In addition, the packets on the right-hand side are substantially amplified in consequence of their single or multiple passage through the gainy slab. Amplification also occurs on the left-hand side, where $x_0$ is negative, although the growth with distance along the negative $x$-axis is not nearly as rapid as that along the positive $x$-axis.

In general, more samples along the $\omega'$-axis and also along the integration contour $C$ in the $k_x$-plane are needed for larger values of $x_0$ and $z_0$. Similarly, more computational resources must be summoned for larger gain coefficients and/or thicker slabs, which cause the $k_x$-plane integration contour $C$ to move further away from the $k_x'$-axis. Given sufficient computing power, one could also contemplate more realistic representations of both passive and active media by introducing additional Lorentz oscillators into the models of the dielectric functions used for media 1, 2, and 3 in accordance with Eq.(9).

**10. Concluding remarks**. This paper has described a systematic approach to computing the Fresnel reflection and transmission of a finite-duration, finite-spatial-footprint wavepacket arriving within a nearly transparent incidence medium at two kinds of interfaces. In the first case, the interface is with a semi-infinite gain medium. In the second case, the interface is with a gain medium of finite-thickness, which is followed by another nearly transparent semi-infinite medium. The crucial step in both cases is to deform the integration path away from the real $k_x'$-axis in the complex $k_x$-plane in such a way as to avoid crossing the relevant branch-cuts and also to eliminate all the poles and singularities from the upper-half of the complex $\omega$-plane. In conjunction with a proper specification of the branch-cuts, this choice of the integration contour in the $k_x$-plane not only ensures the causality of the reflected and transmitted wavepackets, but also leaves no ambiguity regarding the correct signs of the square root expressions that define the $k_z$ components of the various $k$-vectors.

In our numerical simulations, we used a single-oscillator Lorentz model to represent the dielectric functions $\varepsilon(\omega)$ of the incidence, transmittance, and gain media. However, there should be no limitations in principle on the number of such oscillators that could be used to represent each medium. More realistic simulations, especially those involving incidence and transmittance media with reasonably large refractive indices over a broad range of frequencies, would require more than one Lorentz oscillator.

We also avoided the complications arising from material nonlinearities, including gain saturation.[16] It is well known that the field amplitudes inside a gain medium cannot grow indefinitely, and that the degree of population inversion — itself determined by the pump power and by the inherent properties of the material medium — limits the range over which the complex refractive index $n_2(\omega) = n_2' + \mathrm{i}n_2''$ could be considered to be independent of the internal $E$-field amplitudes. Our linear treatment of the gain medium (i.e., treatment in accordance with the so-called "small-signal model") is thus reliable only up until the point in time when gain saturation begins to show its inexorable effects.

Although we chose in this paper to deform the integration contour in the $k_x$-plane while keeping the inverse Fourier integral over the real $\omega'$-axis of the $\omega$-plane, we could instead have invoked similar lines of reasoning (and more or less the same procedural steps) to deform the $\omega$-plane integration contour while keeping the inverse Fourier integral over the real $k_x'$-axis of the $k_x$-plane. The arguments based on causality, the analyticity of the inverse Fourier integrands, and the role of the branch-points and branch-cuts of $k_{1z}$, $k_{2z}$, and $k_{3z}$ would then have led to an alternative, albeit equally valid, formulation; for a more detailed discussion, see Appendix B. In this alternative approach, the deformed integration contour would be entirely in the upper half of the $\omega$-plane and, if desired, it could be constructed to exhibit even symmetry with respect to the $\omega''$-axis (in contrast to the odd symmetry of the $k_x$-plane contour around the $k_x''$-axis).

Finally, let us emphasize the substantial freedom that is available to us in choosing the $k_x$-plane integration contour $C$, so long as the contour does not cross the relevant pole trajectories and/or the branch-cuts. In general, of course, the constraints imposed on the contour $C$ for the semi-infinite gain medium of Fig.1(a) differ from those for the finite-thickness gainy slab of Fig.1(b). As such, the contours chosen for these two cases need not be identical, although, in our reported numerical simulations, we could and did choose the same contour $C$ for both cases. Moreover, our material parameters for the dielectric function $\varepsilon_2(\omega)$ and the thickness $d$ of the gainy slab were deliberately chosen to bring about a fairly small deviation of the contour $C$ away from the real $k_x'$-axis. A larger gain coefficient and/or a thicker slab would cause the contour $C$ to depart further from the $k_x'$-axis, which would then demand higher numerical accuracy and longer computation times.

Our ability to deform the integration path away from the $k_x'$-axis reaches its natural limit when the singularities begin to pass through the imaginary $k_x''$-axis. (Recalling that such singularities arrive in pairs, we note that a singularity on the positive $k_x''$-axis always accompanies one on the negative $k_x''$-axis.) Under such circumstances, construction of a $k_x$-plane contour $C$ will no longer be feasible, and one must revert to integration along the real $k_x'$-axis in conjunction with a deformed $\omega$-plane integration contour — a contour that moves away from the $\omega'$-axis into the upper-half $\omega$-plane, as described in Appendix B. Failure to construct a $k_x$-plane contour $C$ heralds the arrival of so-called absolute instabilities (e.g., onset of lasing), in contrast to "convective" instabilities, which have been the concern of the present paper.[12,14,15]

## Appendix A
## Constructing a realistic incident wavepacket at the front facet of the gain medium

In Sec.8, we described a simple model for the compact wavepacket that arrives at the interface between media 1 and 2. A more realistic description of the incident packet requires that the incidence medium 1 be modeled as a prism whose slanted facet makes an angle $\theta$ with the $x$-axis, as depicted in Fig. A1. The finite-width, finite-duration wavepacket now arrives at the slanted facet, which coincides with the $x'y'$-plane of the tilted $x'y'z'$ coordinate system. The incident $E$-field amplitude in the $x'y'$-plane at $z' = 0$ is given by

$$\boldsymbol{E}^{(\text{inc})}(x', t) = f(x'/W)\, g(t/T) \cos(\omega_c t)\, \hat{\boldsymbol{y}}. \tag{A1}$$

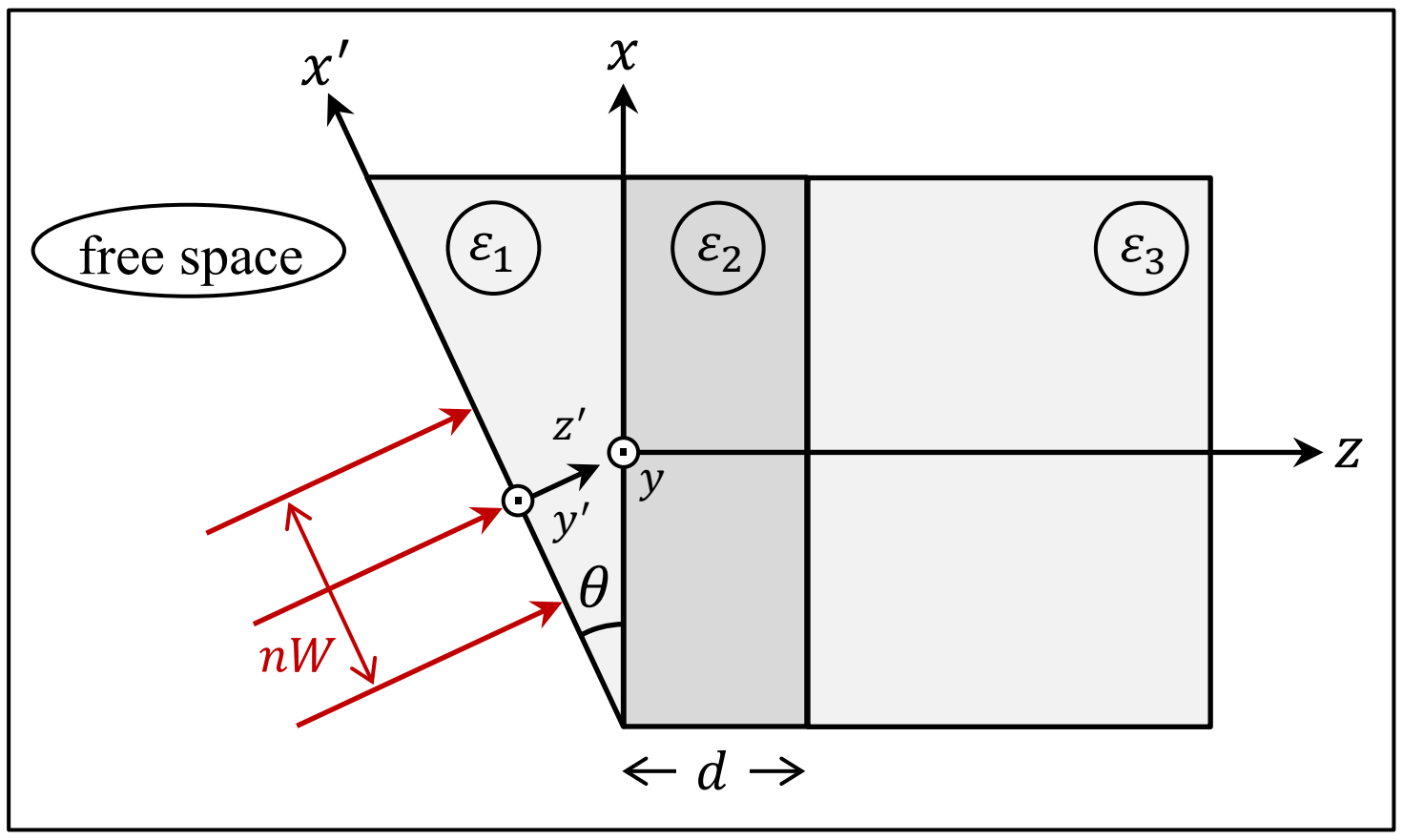

**Fig. A1**. A finite-width, finite-duration wavepacket arrives at normal incidence at the slanted facet of a prism whose dielectric function is specified as $\varepsilon_1(\omega)$. The apex angle of the prism is $\theta$, and the distance that the central incident ray must travel along the $z'$-axis to reach the origin of the $xyz$ coordinate system is $\zeta_0$. The point $(x, z) = (x, 0)$ located at the interface between the incidence medium 1 and gain medium 2 coincides with $(x', z') = (x\cos\theta, \zeta_0 + x\sin\theta)$ in the $x'y'z'$ coordinate system.

Assuming the Fourier transforms of the spatial and temporal profiles of the wavepacket are given by $\tilde{f}(k'_x) = W\,\mathrm{sinc}^n(Wk'_x/2\pi)$ and $\tilde{g}(\omega) = T\,\mathrm{sinc}^m(T\omega/2\pi)$, where $m$ and $n$ are small positive integers, the Fourier transform of the incident $E$-field is readily found to be

$$\tilde{E}_y^{(\mathrm{inc})}(k'_x,\omega) = ½\tilde{f}(k'_x)[\tilde{g}(\omega-\omega_c)+\tilde{g}(\omega+\omega_c)]. \quad \text{(A2)}$$

At the entrance facet of the prism, where $z'=0$, we have $k'_{0z} = (\omega/c)\sqrt{1-(ck'_x/\omega)^2}$ in free space and $k'_{1z} = (\omega/c)\sqrt{\varepsilon_1(\omega)-(ck'_x/\omega)^2}$ in medium 1. In the case of $y$-polarized incident light, the Fresnel transmission coefficient at the prism's slanted facet is given by

$$\tau'(k'_x,\omega) = 2k'_{0z}/(k'_{0z}+k'_{1z}). \quad \text{(A3)}$$

Given that $k'_x$ is real and that $\omega$ is real and positive, one may simplify the computation by choosing the sign of the square roots so that $k'_{0z}$ and $k'_{1z}$ are in $Q_1$ of the complex plane; in other words, there is no need to resort to properly constructed branch-cuts (i.e., straight vertical lines in $Q_1$ and $Q_3$ of the $k_x$-plane). With reference to Fig.A1 and noting that, at the interface between media 1 and 2, $x' = x\cos\theta$ and $z' = \zeta_0 + x\sin\theta$, we now invoke Eqs.(A2) and (A3) to express the incident $E$-field arriving at the interfacial $xy$-plane located at $z=0$ as the following inverse Fourier transform integral:

$$\tilde{E}_y^{(\mathrm{inc})}(x,t) = (2\pi^2)^{-1}\mathrm{Re}\int_{\omega=0}^{\infty}\mathrm{d}\omega e^{-\mathrm{i}\omega t}$$
$$\times\int_{k'_x=-\infty}^{\infty}\tau'(k'_x,\omega)\tilde{E}_y^{(\mathrm{inc})}(k'_x,\omega)e^{\mathrm{i}k'_x x\cos\theta+\mathrm{i}k'_{1z}(\zeta_0+x\sin\theta)}\mathrm{d}k'_x. \quad \text{(A4)}$$

In this equation, the inner integral over $k'_x$ must be evaluated as a function of $x$ for each real and positive value of $\omega$. In what follows, this function will be referred to as $E_y^{(\mathrm{inc})}(x,\omega)$, that is,

$$E_y^{(\mathrm{inc})}(x,\omega) = (2\pi)^{-1}\int_{k'_x=-\infty}^{\infty}\tau'(k'_x,\omega)\tilde{E}_y^{(\mathrm{inc})}(k'_x,\omega)e^{\mathrm{i}k'_x x\cos\theta+\mathrm{i}k'_{1z}(\zeta_0+x\sin\theta)}\mathrm{d}k'_x. \quad \text{(A5)}$$

Note that $\zeta_0$, the distance between the origins of the $xyz$ and $x'y'z'$ coordinates must be chosen such that $\zeta_0 + x\sin\theta$ is non-negative for all values of $x$ in the interval $[x_{\min}, x_{\max}]$; that is, $\zeta_0 \geq |x_{\min}|\sin\theta$. Also, the integral in Eq.(A5) is evaluated over the real $k'_x$ axis. Having computed $E_y^{(\mathrm{inc})}(x,\omega)$ for all real and positive frequencies $\omega$, we proceed to find its Fourier transform over $x$, namely,

$$\tilde{E}_y^{(\mathrm{inc})}(k_x,\omega) = \int_{-\infty}^{\infty}E_y^{(\mathrm{inc})}(x,\omega)e^{-\mathrm{i}k_x x}\mathrm{d}x. \quad \text{(A6)}$$

This is the function that now substitutes for $\tilde{E}_y^{(\mathrm{inc})}(k_x,\omega')$ in Eqs.(7) and (8).

## Appendix B

### Deforming the integration path in the complex $\omega$-plane

A glance at Figs.8 and 9 reveals that the $k_x$-plane trajectories of poles and branch-points of a given system generally move upward when the temporal frequency $\omega$, starting at a positive real value $\omega'$, acquires a positive imaginary part $\omega''$ by moving up, parallel to the imaginary axis of the $\omega$-plane. Not shown in Figs.8 and 9 are the simultaneous happenings on the left half of the $k_x$-plane, where, due to the inherent odd symmetry, all the pole and branch-point trajectories move downward. It is thus seen that the real $k'_x$-axis can be cleared of all the singularities of the inverse Fourier integrands of Eqs.(7) and (8) if the $\omega$-plane integration contour is sufficiently moved away from the real $\omega'$-axis and into the upper-half of the $\omega$-plane.

As a simple example, note that the Fourier transformation of the incident packet at $z = 0$ can be done on any straight-line $\omega = \omega' + \mathrm{i}\Omega_0$ that is parallel to the $\omega'$-axis, provided that $\Omega_0 \geq 0$. (The vanishing of the incident packet for $t < 0$ guarantees the existence of its Fourier transform for any value of $\omega'' \geq 0$.) In this way, one can proceed to solve Maxwell's equations for individual plane-waves in media 1, 2, and 3, match the boundary conditions at $z = 0$ and $z = d$, and obtain the usual Fresnel reflection and transmission coefficients, $\rho(k'_x, \omega)$ and $\tau(k'_x, \omega)$, with the tacit assumption that $\omega$ is an arbitrary point on the straight line parallel to and above the $\omega'$-axis.[12]

At this point in the analysis, the existence of branch-points for $k_{2z}$ (i.e., $k_z$ in medium 2) and/or poles associated with $\rho$ and $\tau$ in the upper-half $\omega$-plane imposes a lower bound on $\Omega_0$ that ensures the satisfaction of the all-important causality requirement.[13-15] (Causality decrees that the reflected and transmitted waves cannot reach the point $(x, y, z)$ prior to $t = |z|/c$.) Causality also fixes the signs of $k_{1z}$, $k_{2z}$, and $k_{3z}$ so that, referring to Eq.(6), there will be no ambiguity as to which one of the $\pm$ signs should be picked at any given point $(k'_x, \omega' + \mathrm{i}\Omega_0)$ in the Fourier domain. In this way, the reflected and transmitted EM fields at the observation point $(x_0, z_0, t)$ can be computed via a 2D inverse Fourier transformation, first over the $k'_x$-axis, and then along the straight-line $\omega = \omega' + \mathrm{i}\Omega_0$ in the $\omega$-plane.

As we have argued in this paper, under certain circumstances, the Fourier transforms can be rearranged in such a way that the transform in the $\omega$-plane returns to the real $\omega'$-axis at the expense of carrying out the Fourier integral in the $k_x$-plane over a properly deformed contour — as opposed to over the real $k'_x$-axis. In accordance with the arguments advanced in Sec.2, the deformed integration contour in the $k_x$-plane is chosen such that the associated singularities will disappear from the upper-half $\omega$-plane, thereby clearing the way for the integration path $\omega = \omega' + \mathrm{i}\Omega_0$ to return to the $\omega'$-axis. The mathematical basis for these assertions, of course, continues to be the Cauchy-Goursat theorem of complex analysis.[9-11] It is worth emphasizing once again that the choice of the integration path, be it the straight-line $\omega = \omega' + \mathrm{i}\Omega_0$ in the $\omega$-plane or a properly deformed contour in the $k_x$-plane, is dictated by the analyticity of the functions involved and by the requirement of causality. These constraints also automatically fix the signs of $k_{1z}$, $k_{2z}$, and $k_{3z}$ *without* resort to any kind of "commonsense" physical argument.

**Acknowledgement**. The authors are grateful to Tobias Mansuripur and Johannes Skaar for many helpful discussions.

**Conflict of interest statement**. On behalf of all authors, the corresponding author states that there is no conflict of interest.